%% file: Hardware_Journal_2024.tex
\documentclass[lettersize,journal]{IEEEtran}
\usepackage{amsmath,amsfonts}
\usepackage{algorithmic}
\usepackage{algorithm}
\usepackage{array}
\usepackage{textcomp}
\usepackage{stfloats}
\usepackage{url}
\usepackage{color}
\usepackage{verbatim}
\usepackage{graphicx}
\usepackage{cite}

\usepackage{subfigure}
\begin{document}

\title{DNN based Two-stage Compensation Algorithm for THz Hybrid Beamforming with imperfect Hardware}

\author{Wenqi Zhao, Chong Han,~\IEEEmembership{Senior Member,~IEEE}, Ho-Jin Song,~\IEEEmembership{Fellow,~IEEE}, 
and Emil Bj{\"o}rnson,~\IEEEmembership{Fellow,~IEEE}
\thanks{This paper was presented in part at IEEE ICC, Jun. 2024~\cite{Zhao-2024-Dynamic}.
    
    W.-Q. Zhao is with the Terahertz Wireless Communications (TWC) Laboratory, Shanghai Jiao Tong University, Shanghai 200240,
    China (e-mail: wenqi.zhao@sjtu.edu.cn).

    C. Han is with Terahertz Wireless Communications (TWC) Laboratory,
    and with the Department of Electronic Engineering and Cooperative Medianet
    Innovation Center (CMIC), Shanghai Jiao Tong University, Shanghai 200240,
    China (e-mail: chong.han@sjtu.edu.cn).

    H. Song is with the School of Electrical Engineering, Pohang University of Science and Technology, Pohang 37673, South Korea (e-mail: hojin.song@postech.ac.kr).

    E. Bj{\"o}rnson is with the Department of Computer Science, KTH Royal Institute of Technology, Stockholm, Sweden (email: emilbjo@kth.se).}}



\maketitle
	\thispagestyle{empty}

\begin{abstract}
Terahertz (THz) communication is envisioned as a key technology for 6G and beyond wireless systems owing to its multi-GHz bandwidth. 
To maintain the same aperture area and the same link budget as the lower frequencies,
ultra-massive multi-input and multi-output (UM-MIMO) with hybrid beamforming is promising. Nevertheless, the hardware imperfections particularly at THz frequencies, can degrade spectral efficiency and lead to a high symbol error rate (SER), which is often overlooked yet imperative to address in practical THz communication systems. In this paper, the hybrid beamforming is investigated for THz UM-MIMO systems accounting for comprehensive hardware imperfections, including DAC and ADC quantization errors, in-phase and quadrature imbalance (IQ imbalance), phase noise, amplitude and phase error of imperfect phase shifters and power amplifier (PA) nonlinearity. Then, a two-stage hardware imperfection compensation algorithm is proposed. A deep neural network (DNN) is developed in the first stage to represent the combined hardware imperfections, while in the second stage, the digital precoder in the transmitter (Tx) or the combiner in the receiver (Rx) is designed using NN to effectively compensate for these imperfections. Furthermore, to balance the performance and network complexity, three slimming methods including pruning, parameter sharing, and removing parts of the network are proposed and combined to slim the DNN in the first stage. Numerical results show that the Tx compensation can perform better than the Rx compensation. Additionally, using the combined slimming methods can reduce parameters by 97.2\% and running time by 39.2\% while maintaining nearly the same performance in both uncoded and coded systems.
\end{abstract}


\section{Introduction}
\IEEEPARstart{O}{wing} to the demand of ultra-broad multi-GHz bandwidth, Terahertz (THz) wireless communication has been envisioned as one of the key technologies in the sixth generation (6G) and beyond communication systems~\cite{I-2014-Terahertz}. To maintain the same aperture area and the same link budget as the lower frequencies, ultra-massive multi-input and multi-out (UM-MIMO) is needed which can support thousands of antennas at the transceivers. In THz UM-MIMO systems, the hybrid beamforming is investigated as a promising technology~\cite{Dai-2016-Hybrid},~\cite{Yan-2020-A} which is sufficient to make use of THz channels with few channel paths. Moreover, hybrid beamforming is less expensive to implement than full digital beamforming while achieving high spectral efficiency~\cite{Sohrabi-2016-Hybrid}.


However, one of the main challenges in implementing THz UM-MIMO systems is the exacerbated impact of hardware imperfections, primarily due to the high carrier frequency of THz, which increases sensitivity to phase noise. Additionally, the large bandwidth and high frequency of THz communication significantly raise the complexity and cost of baseband and RF hardware manufacturing~\cite{Song-2022-Terahertz,Huang-2024-Received}.
Some hardware imperfections that have minimal impact in the low-frequency band have caused significant effects in higher frequency band such as millimeter waves and terahertz, making it essential for us to pay more attention to these imperfections in system design and practical implementation. Specifically, in the baseband of both the transmitter~(Tx) and receiver~(Rx), digital-to-analog converters (DACs) and analog-to-digital converters (ADCs) can introduce quantization errors~\cite{zhang-2016-On,Orhan-2015-Low}. In RF chains, in-phase and quadrature imbalance (IQ imbalance)\cite{Lee-2024-248-GHz,Mahendra-2022-Downlink,Mahendra-2020-Transmitter} as well as phase noise\cite{Chatelier-2022-on-PN,Forsch-2022-phase} caused by imperfect oscillators lead to severe nonlinearity. In antenna arrays, imperfect phase shifters~\cite{Muller-2017-A,Wu-2022-A-220-GHz} can introduce amplitude and phase errors to the passing signals, while nonlinear power amplifiers (PAs)~\cite{Choi-2023-275-GHz,Tervo-2023-Parametrization,Moghadam-2018-on,Abdelaziz-2018-digital-predistortion} exhibit increasingly nonlinear characteristics as the signal power increases. These hardware imperfections significantly impact the performance of THz UM-MIMO systems and remain to be solved.

\subsection{Related Work}
In the literature, many techniques have been considered to mitigate the impact of hardware imperfections.
To handle the quantization error of imperfect DACs and ADCs, the authors in \cite{zhang-2018-On-low-resolution} proposed a mixed-ADC architecture combining high- and low-resolution ADCs to enhance system performance in millimeter-wave systems, while the author in \cite{Verenzuela-2017-Per-antenna} demonstrates that using equal-resolution ADCs across the antenna array outperforms the mixed-ADC architecture, achieving higher spectral efficiency and lower power consumption.
To address the IQ imbalance, in~\cite{Mahendra-2022-Downlink}, the authors considered the downlink multi-user hybrid beamforming systems with receiver IQ imbalance and propose an Rx-IQ imbalance pre-compensation algorithm that nearly matches the performance of an IQ imbalance-free system. However, the Tx IQ imbalance is not considered and the evaluation is limited to mmWave frequencies.
As for phase noise,~\cite{Chatelier-2022-on-PN} investigated the effects of phase noise on the array factor and beamforming gain of massive MIMO mmWave systems. Additionally,~\cite{Forsch-2022-phase} analyzed phase noise in THz communications and verified that the Gaussian phase noise model can accurately approximate phase noise in the THz band with A free-running Voltage-Controlled Oscillator. To solve the imperfect phase shifters, in~\cite{yang-2019-Hardware-Constrained}, the lens-based antenna arrays were proposed to substitute the phased antenna arrays which continue to face manufacturing imperfectness and are not appropriate in THz UM-MIMO. As for the nonlinear PAs, the author in~\cite{Moghadam-2018-on} studied the performance of mmWave hybrid beamforming scheme in the presence of nonlinear PAs and investigated the trade-off between spectral and energy efficiency. In~\cite{Abdelaziz-2018-digital-predistortion}, a novel digital pre-distortion (DPD) processing was proposed to compensate the nonlinear PAs in sub-array hybrid beamforming architecture which needs to obtain the signal transmitted from antennas and down-convert the signal to baseband. This scheme needs extra RF chains and ADCs in the Tx, which significantly increases both hardware complexity and cost, especially in the THz band. Most of the aforementioned methods in the literature only focus on and deal with a single type of hardware imperfection.

Some research takes multiple hardware imperfections into consideration~\cite{Björnson-2014-Massive,Björnson-2019-Hardware,Javed-2019-Multiple,Zhang-2021-Cell-Free,Huang-2024-Received,Zhang-2014-On}. In~\cite{Björnson-2014-Massive}, the authors considered a system model that incorporates general transceiver hardware imperfections at the Tx and Rx. 
In~\cite{Björnson-2019-Hardware}, the authors investigated how distortion caused by the hardware imperfections impacts the uplink spectral efficiency. In~\cite{Huang-2024-Received}, a simple received signal model was proposed for a wideband THz system with nonlinearity and other hardware imperfections to enable low-complexity mitigation. However, this model only described the hardware imperfections in the single antenna system rather than a MIMO system. In~\cite{Zhang-2021-Cell-Free}, the author considered the phase drifts and distortion noise in the uplink MIMO system and designed a power optimization algorithm to maximize the sum rate.
Moreover, machine learning and deep learning have been considered to mitigate the combined impact of hardware imperfections~\cite{Qing-2022-Joint,Jaraut-2018-Composite,Wang-2019-Augmented,Gao-2024-Signal}. 
In~\cite{Qing-2022-Joint}, a data-dependent superimposed training scheme was used to mitigate the symbol misidentification caused by hardware imperfections. An NN-based digital pre-distortion was proposed in ~\cite{Jaraut-2018-Composite} to compensate for crosstalk, PA nonlinearity, and IQ imbalance. In ~\cite{Gao-2024-Signal}, a deep neural network (DNN) technique was proposed to address the combined nonlinearity of power amplifiers (PAs) and IQ imbalance in uplink multi-user mmWave systems. Additionally, the aforementioned compensation schemes do not account for the hybrid beamforming architecture in THz UM-MIMO systems. Therefore, it is crucial to consider comprehensive hardware imperfections in THz hybrid beamforming and to develop more effective compensation methods with low complexity for these hardware imperfections. 

\subsection{Contributions}
In this work, we investigate the hardware imperfections in THz hybrid beamforming UM-MIMO systems, where the transmitter suffers from combined distortions of imperfect DACs, IQ imbalance, phase noise, imperfect phase shifters, and nonlinear PAs and the receiver suffers from combined distortions of imperfect phase shifters, IQ imbalance, phase noise and imperfect ADCs. To mitigate the hardware imperfections and reduce the symbol error rate (SER), we propose a two-stage hardware imperfection compensation algorithm based on DNN. Furthermore, to balance the compensation performance and network complexity, we also propose three slimming methods and combine them to reduce the complexity of DNN in the first stage. The contributions of this paper are summarized as follows.
\begin{itemize}

\item[$\bullet$] \textbf{We present the hardware imperfection models in the THz hybrid beamforming system.} Compared with the ideal MIMO system, the hybrid beamforming accounting for comprehensive hardware imperfections in the transceiver is analyzed. Then, We present all hardware imperfect models individually and give a model with the combined effect of all hardware imperfections.
\item[$\bullet$] \textbf{We propose a two-stage hardware imperfection compensation algorithm for THz hybrid beamforming based on deep neural networks (DNN).} In the first stage, we treat the combined hardware imperfections as black boxes and develop a deep neural network (DNN), which is composed of multiple sub-neural networks (sub-NNs) to represent their features based on the signal transmission process. In the second stage, utilizing the DNN developed in the first stage, we design the digital precoder at the Tx or the combiner at the Rx using neural networks (NN) to effectively compensate for the hardware imperfections.
\item[$\bullet$] \textbf{We develop three methods to slim the DNN developed in the first stage.} 
The first method involves pruning the DNN by reducing the number of neurons in the hidden layers of each sub-NN. This approach decreases the number of parameters with minimal performance degradation. The second method involves sharing parameters among sub-NNs that represent similar hardware imperfections. This strategy not only preserves relatively good performance but also significantly reduces both the number of neural network parameters and the runtime. The third method is applicable when the transmit power does not cause PA nonlinearity, allowing for the removal of certain parts of the DNN as the hardware imperfections become simpler. By combining these slimming methods, we can achieve a further reduction in computational complexity while preserving a good performance.
\item[$\bullet$] \textbf{We analyze the loss caused by hardware imperfections and evaluate the performance of the two-stage compensation algorithm and slimming methods in both uncoded and coded systems.} First, we present the system parameters and analyze hardware imperfections in terms of SER and signal constellation. We then evaluate the performance of the proposed compensation and slimming methods. The results demonstrate that the compensation algorithm can effectively mitigate hardware imperfections, reduce the SER, and the performance can be maintained with appropriate slimming using the three methods and their combinations in both uncoded and coded systems. 
\end{itemize}

The remainder of the paper is organized as follows. In Sec.~\ref{II}, the system models of the THz MIMO system with ideal and non-ideal hardware are investigated. In Sec.~\ref{III}, we introduce the hardware imperfection models and the expression of received signals effected by combined hardware imperfections. The problem formulation of hardware imperfection compensation is investigated In Sec.~\ref{IV}.
In Sec. \ref{V} and Sec.~\ref{VI}, the compensation algorithm and neural network slimming methods are proposed, respectively. In Sec.~\ref{VII}, we analyze the loss caused by hardware imperfections and evaluate the performance of the proposed methods. Finally, the conclusion is drawn in Sec.~\ref{VIII}.

\textbf{Notation:}
$a$ is a scalar and $\mathbf{a}$ is a vector. $\mathbf{A}$ denotes a matrix and $\mathbf{A}(m,n)$ is the element at $m^{th}$ row and $n^{th}$ column in $\mathbf{A}$.
$\mathbb{C}^{M\times N}$ represents $M \times N $-dimensional complex-valued matrices. $\mathbf{I}_N$ defines an N dimensional identity matrix. $(\cdot)^*$,  $(\cdot)^\mathrm{T}$ and $(\cdot)^\mathrm{H}$ refer to conjugate, transpose and conjugate transpose. $\mathbb{E}(\cdot)$ describes the expectation and $\text{diag}(\mathbf{A})$ is keeps only the diagonal entries of $\mathbf{A}$. $\mathrm{Re}(\cdot)$ and $\mathrm{Im}(\cdot)$ refer to the real and imaginary part of a complex number. $|\cdot|$, $||\cdot||$ and $||\cdot||_{\mathrm{F}}$ denote absolute value, 2-norm and Frobenius norm.
\begin{figure*}
    \centering
\includegraphics[scale=0.25]{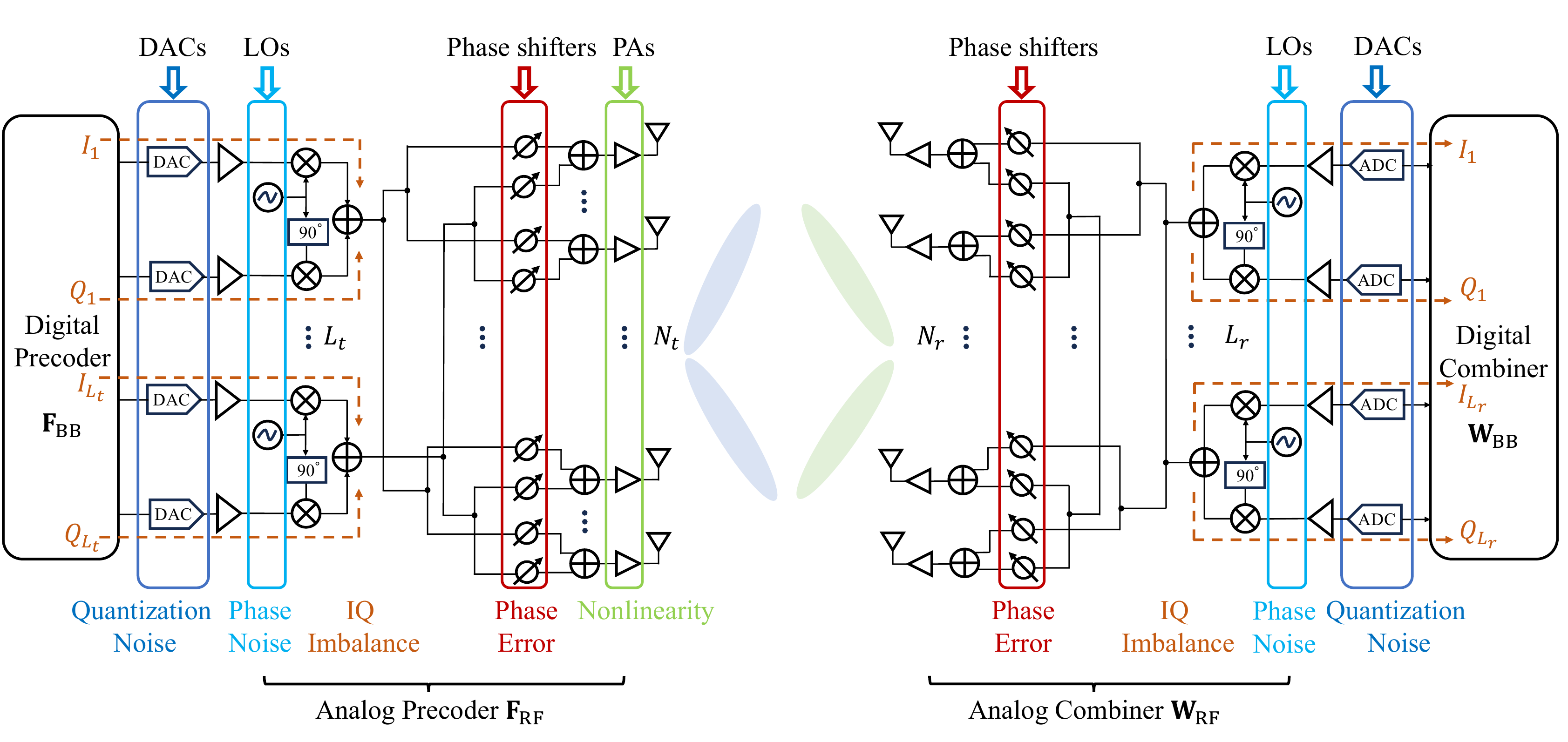}
    \caption{THz UM-MIMO system with imperfect hardware.}
    \label{fig: system model}
    \vspace{-5mm}
\end{figure*}

\section{System Model}\label{II}

In this section, we basically consider a narrow band THz communication system with a coherence bandwidth of 100~MHz to 1~GHz\cite{Han-2022-Terahertz}. Firstly, we introduce an ideal MIMO system and regard it as a benchmark to illustrate the non-ideal MIMO system with imperfect hardware.

\textit{1) Ideal MIMO System:}
As shown in Fig.~\ref{fig: system model}, we consider a generalized THz UM-MIMO communication system deploying a full-connected hybrid beamforming architecture at both the Tx and Rx sides, respectively. 
To transmit $N_s$ data streams, there are $N_t$ transmit antennas and $N_r$ receive antennas with $L_t$ and $L_r$ RF chains in the Tx and Rx, respectively, which satisfy $N_s\leq L_t < N_t$ and $N_s \leq L_r < N_r$.
We use $\mathbf{s} \in \mathbb{C}^{N_s \times T}$ as the transmitted pilot signal, where $T$ denotes the length of the pilot. At the Tx, the pilot passes through the digital precoder $\mathbf{F}_{\mathrm{BB}} \in \mathbb{C}^{L_t \times N_s}$ and diagonal input power allocation matrix $\mathbf{P}_{\mathrm{in}}=\mathrm{diag}(p_1, \cdots, p_{L_t}) \in \mathbb{C}^{L_t \times L_t}$. 

Subsequently, the signal is processed by analog precoder $\mathbf{F}_{\mathrm{RF}}\in \mathbb{C}^{N_t \times L_t}$ where each element satisfies the constant module constraint, i.e., $|\mathbf{F}_{\mathrm{RF}}(i,j)|=\frac{1}{\sqrt{N_t}}$. Additionally, to meet the total transmit power constraint, $\mathbf{F}_{\mathrm{BB}}$ should be normalized to satisfy $||\mathbf{F}_{\mathrm{RF}}\mathbf{F}_{\mathrm{BB}}||^2_{\mathrm{F}}=N_s$.
After passing through the analog precoder, the signal at each antenna obtains the same gain $g$ from the ideal linear PAs represented by the gain matrix $\mathbf{G}_{\mathrm{PA}}=\mathrm{diag}(g_1, \cdots, g_{N_t})$ where $g_1=g_{N_t}=g$. Therefore, the transmitted signal $\mathbf{x}_i \in \mathbb{C}^{N_t \times T}$ in the ideal MIMO system can be expressed as
\begin{equation}
    \label{ideal transmitted signal}
\mathbf{x}_i=\mathbf{G}_{\mathrm{PA}}\mathbf{F}_{\mathrm{RF}}\mathbf{P}_{\mathrm{in}}\mathbf{F}_\mathrm{BB}\mathbf{s}
\end{equation}

After passing through the channel $\mathbf{H}\in \mathbb{C}^{N_r \times N_t}$, the analog and digital combiners $\mathbf{W}_{\mathrm{RF}} \in \mathbb{C}^{N_r \times L_r}$ and $\mathbf{W}_{\mathrm{BB}} \in \mathbb{C}^{L_r \times N_s}$ are applied to the received signal at the Rx. Similar to the Tx, the analog combiner also follows the constant module constraint, i.e., $|\mathbf{W}_{\mathrm{RF}}(i,j)|=\frac{1}{\sqrt{N_r}}$ and the digital combiner should be normalized to satisfy $||\mathbf{W}_{\mathrm{BB}}^\mathrm{H}\mathbf{W}_{\mathrm{RF}}^\mathrm{H}||^2_{\mathrm{F}}=N_s$. 
Then, the received signal $\mathbf{y}_{i} \in \mathbb{C}^{N_s \times T}$ in ideal MIMO system is
\begin{equation}
    \label{eq:ideal received signal}
    \mathbf{y}_{i}= \mathbf{W}_{\mathrm{BB}}^\mathrm{H} \mathbf{W}_{\mathrm{RF}}^\mathrm{H}\mathbf{H}\mathbf{x}_i+\mathbf{W}_{\mathrm{BB}}^\mathrm{H} \mathbf{W}_{\mathrm{RF}}^\mathrm{H}\mathbf{n},
\end{equation}
where $\mathbf{n} \sim \mathcal{CN}(0,\sigma^2\mathbf{I}_{N_r})$ represents the circularly symmetric complex Gaussian distributed additive noise vector, with $\sigma^2$ noise power.

\textit{2) Non-ideal MIMO System:}
As shown in Fig.~\ref{fig: system model}, the baseband signals processed by digital precoder $\mathbf{F}_{\mathrm{BB}}$ are separated to the real and imaginary parts to pass through the imperfect DACs in I and Q branches respectively, denoted as $\mathcal{Q}_{\mathrm{DAC}}(\cdot)$, which inevitably introduces quantization noise and distorts the signals. Then the distorted signals in the I and Q branches are up-converted to the THz band by imperfect oscillators.  Unlike ideal oscillators and $90^\circ$ phase shifters, which provide signals with equal amplitude and orthogonal phases for the I and Q branches, the imperfect oscillators and imperfect $90^{\circ}$ phase shifters introduce amplitude and phase errors, denoted by $\mathcal{Q}_{\mathrm{IQt}}(\cdot)$, which are commonly referred to as IQ imbalance. Besides, the imperfect oscillators can also introduce phase noise $\mathbf{\Theta}_{\mathrm{t}}\in \mathbb{C}^{L_t \times L_t}$, resulting in an overall phase shift in both I and Q branches. 

After up-converted to the THz band, the signals are processed by the analog precoder $\mathbf{F}_{\mathrm{RF,e}}\in \mathbb{C}^{N_t\times L_t}$, which is equipped by low-resolution phase shifters with both amplitude and phase errors. Then, the signals are input to the nonlinear PAs denoted as $\mathcal{Q}_{\mathrm{PA}}(\cdot)$ which introduces varying gains and phase shifts depending on the input power.
Consequently, the transmitted signals $\mathbf{x}_e \in \mathbb{C}^{N_t \times T}$ in non-ideal MIMO systems can be expressed as 
\begin{equation}
    \label{eq:imperfect transmitted signal}
    \mathbf{x}_e=\mathcal{Q}_{\mathrm{PA}}(\mathbf{F}_{\mathrm{RF,e}}\mathbf{\Theta}_{\mathrm{t}}\mathcal{Q}_{\mathrm{IQt}}( \mathbf{P}_{\mathrm{in}}\mathcal{Q}_{\mathrm{DAC}}(\mathbf{F}_\mathrm{BB}\mathbf{s}))).
\end{equation}

After passing through the channel $\mathbf{H}$, the received signals is amplified by low noise amplifiers (LNA). Although LNAs exhibit nonlinear characteristics similar to PAs, the significant free-space path loss at THz frequencies results in a considerably low signal power received by each antenna at the receiver, making the LNA’s nonlinear effects negligible \cite{Mao-2012-245-GHz}. Consequently, the nonlinearity of LNAs is not considered in this study. After amplified by LNAs, the analog combiner with imperfect phase shifters $\mathbf{W}_{\mathrm{RF,e}}\in\mathbb{C}^{N_r \times L_r}$ combines the signals to $L_r$ branches. Similar to the Tx, down-conversion at the Rx introduces phase noise $\mathbf{\Theta}_{\mathrm{r}} \in \mathbb{C}^{L_r \times L_r}$ and IQ imbalance $\mathcal{Q}_{\mathrm{IQr}}(\cdot)$ due to the imperfect oscillators. Finally, the signals with quantization noise caused by the imperfect ADCs $\mathcal{Q}_{\mathrm{ADC}}(\cdot)$ are processed by the digital combiner $\mathbf{W}_{\mathrm{BB}}$. Thus, the received signal $\mathbf{y}_e \in \mathbb{C}^{N_s \times T}$ in the non-ideal MIMO system is
\begin{equation}
    \label{eq:imperfect received signal}
    \mathbf{y}_e=\mathbf{W}_\mathrm{BB}^\mathrm{H}\mathcal{Q}_{\mathrm{ADC}}(\mathbf{\Theta}_{\mathrm{r}}\mathcal{Q}_{\mathrm{IQr}}(\mathbf{W}_\mathrm{RF,e}^\mathrm{H}\mathbf{H}\mathbf{x}_e+\mathbf{W}_\mathrm{RF,e}^\mathrm{H}\mathbf{n})).
\end{equation}

In the previous content, we have analyzed the various hardware imperfections in the THz UM-MIMO system. Next, in Sec.\ref{III}, we will present all the models of the hardware imperfections as described in \eqref{eq:imperfect received signal}.

\begin{figure*}[ht]\label{fig:constellation}
    \centering
    \subfigure[Ideal hardware.]{\includegraphics[width=1.4in]{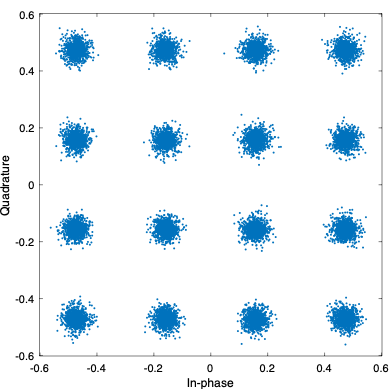}}
    \subfigure[DACs and ADCs (expansion).]{\includegraphics[width=1.4in]{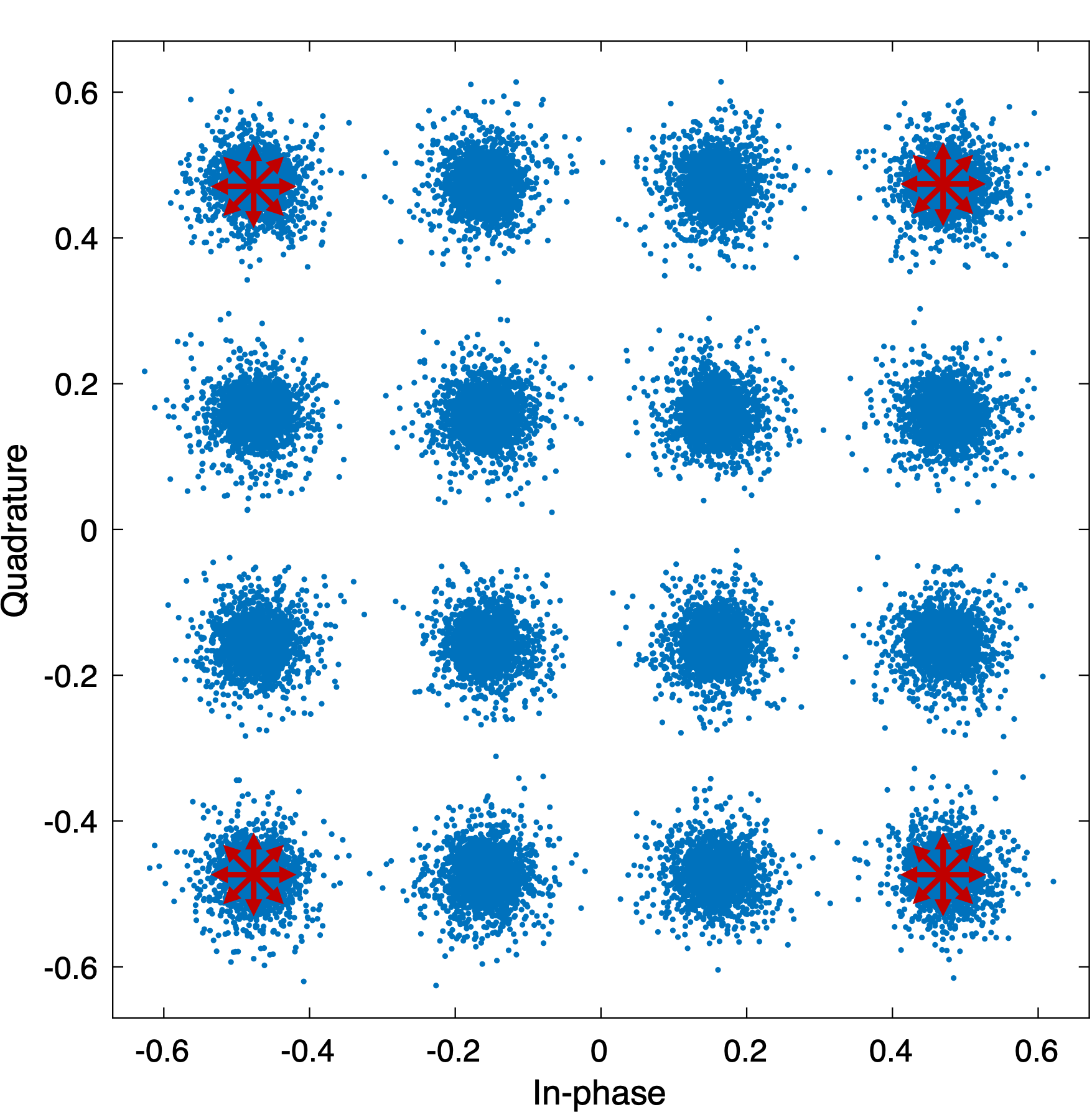}}
    \subfigure[Phase shifters (expansion).]{\includegraphics[width=1.4in]{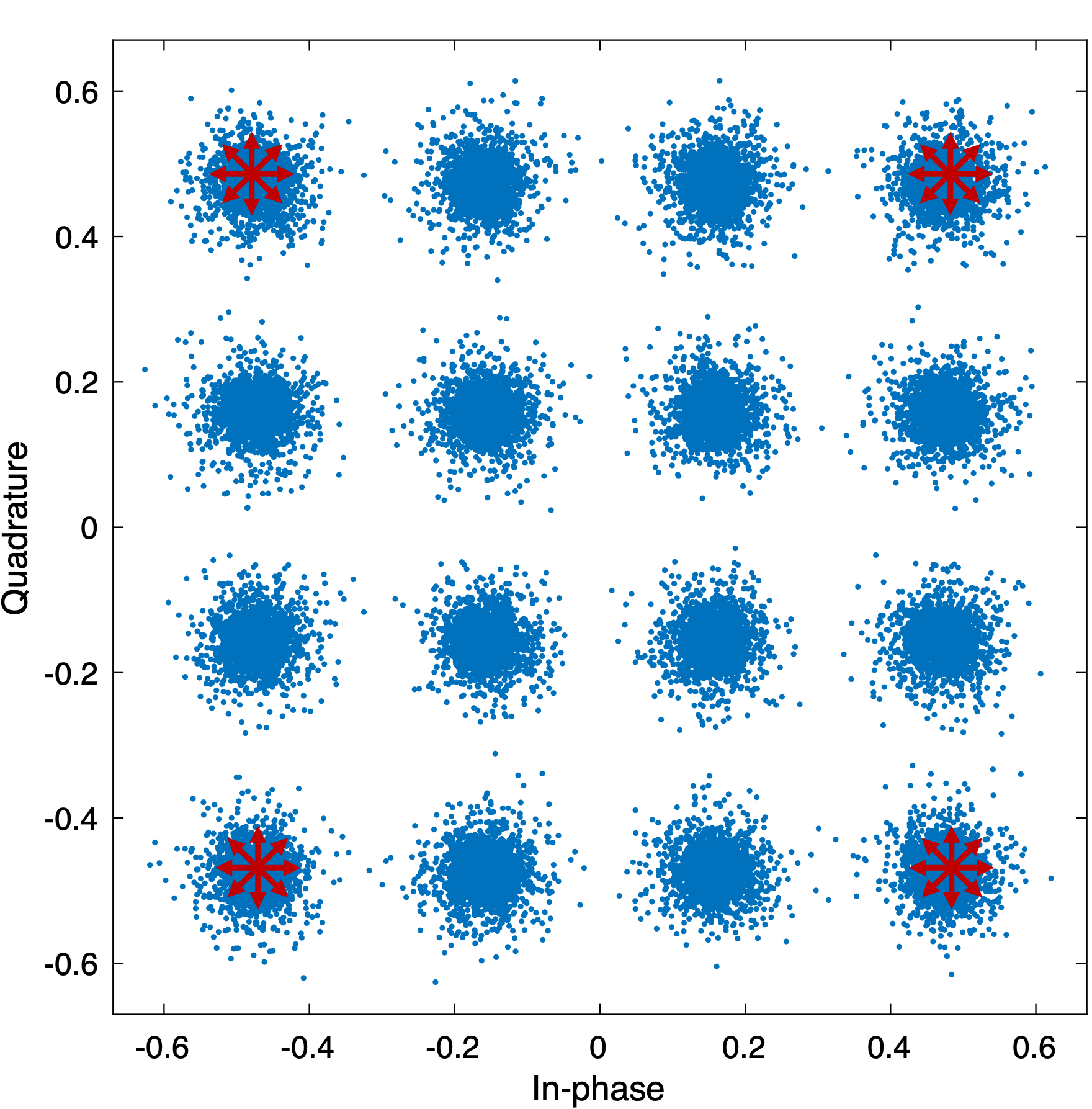}} 
    \subfigure[IQ imbalance (displacement).]{\includegraphics[width=1.4in]{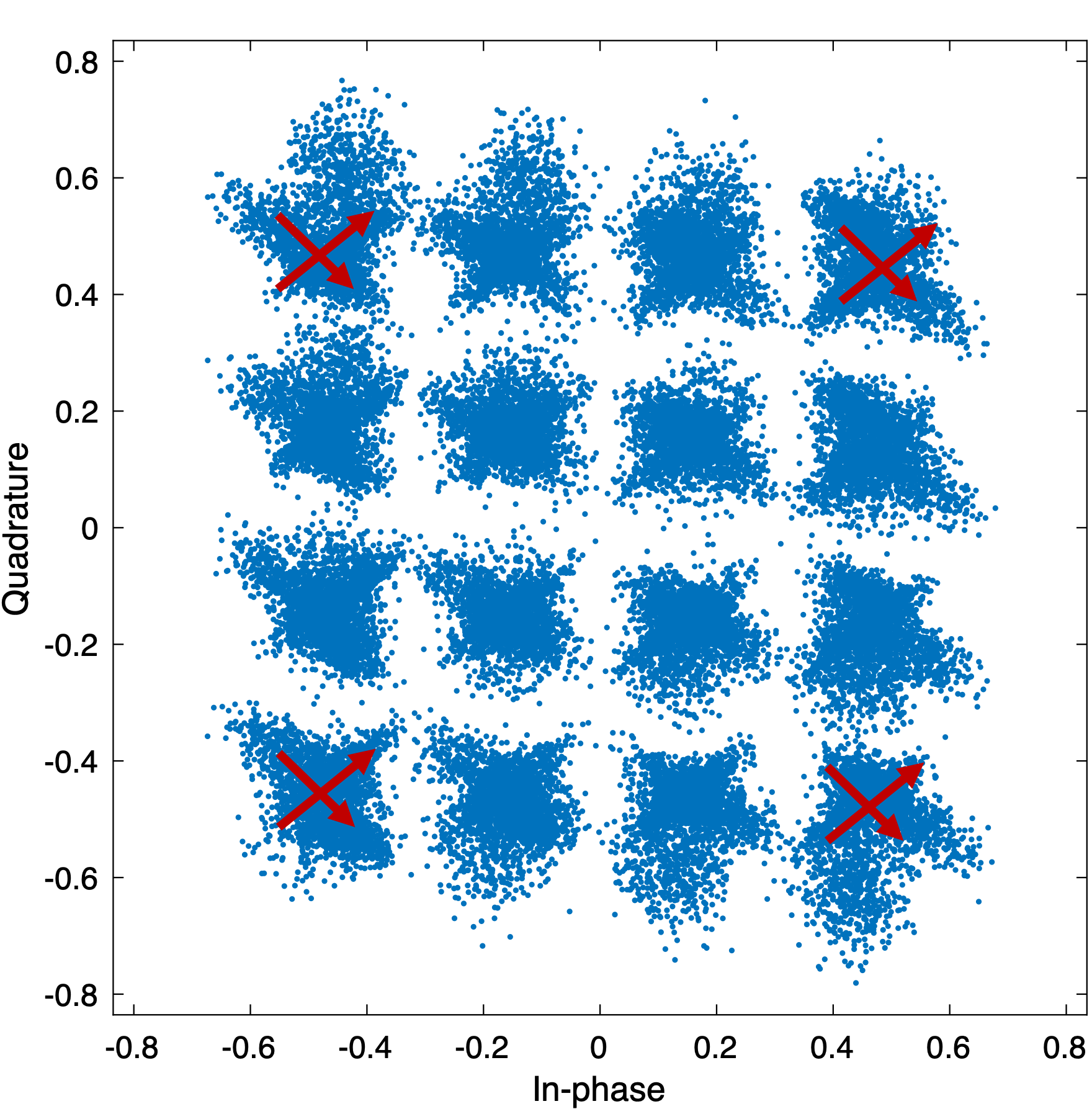}}\\
    \subfigure[Phase noise (rotation).]{\includegraphics[width=1.4in]{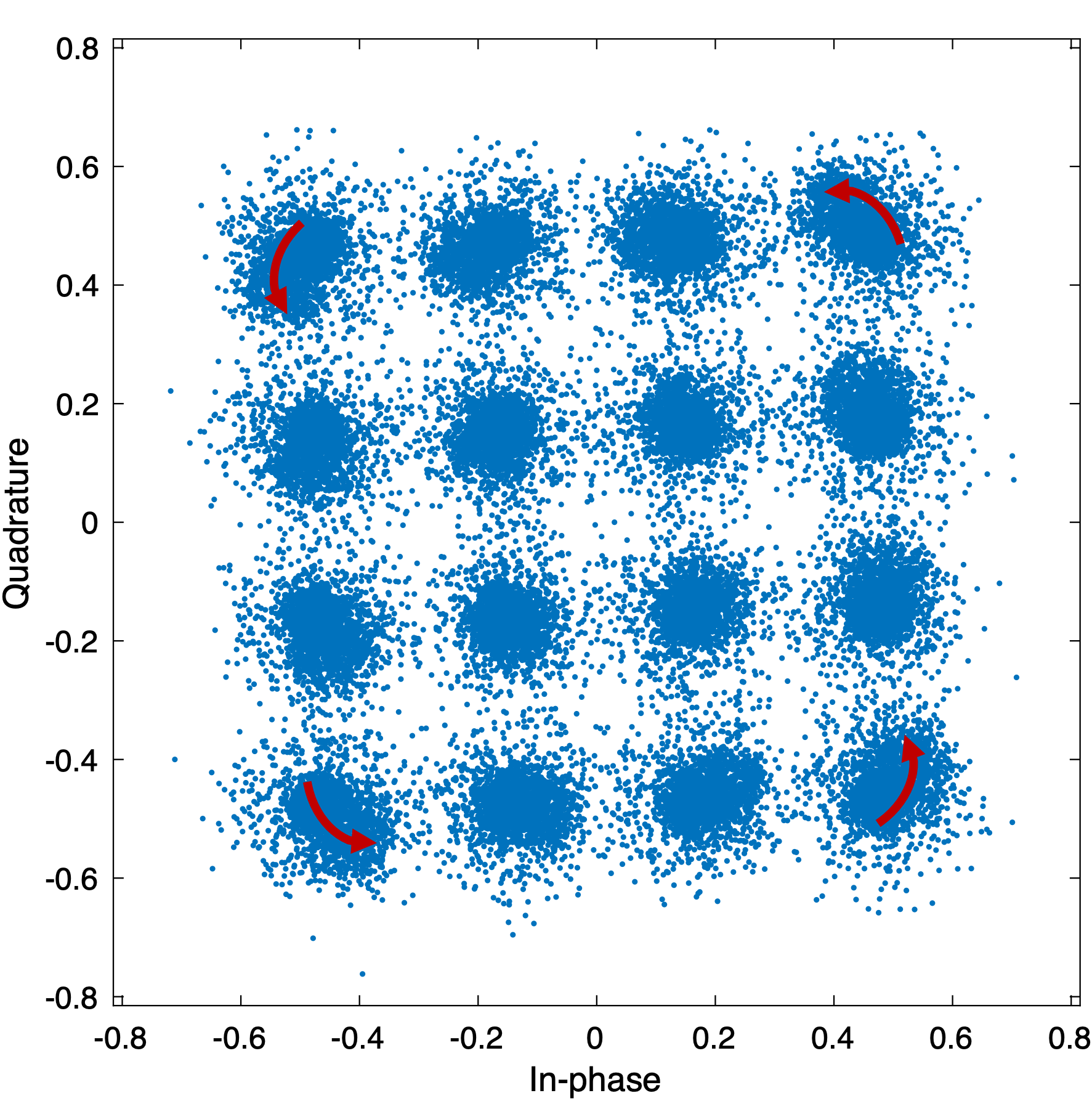}}
    \subfigure[Nonlinear PAs (rotation).]{\includegraphics[width=1.4in]{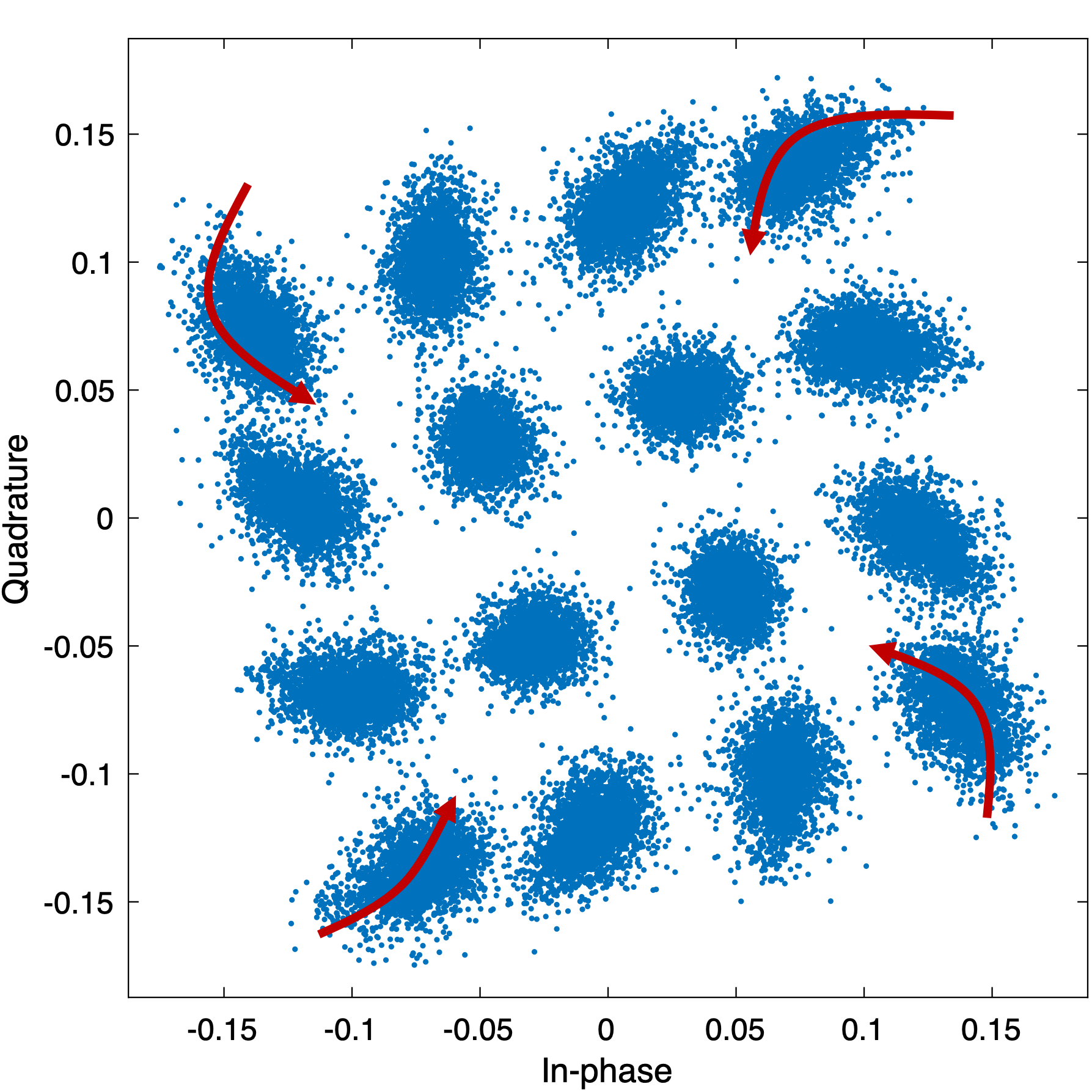}}
     \subfigure[Non-ideal hardware.]{\includegraphics[width=1.62in]{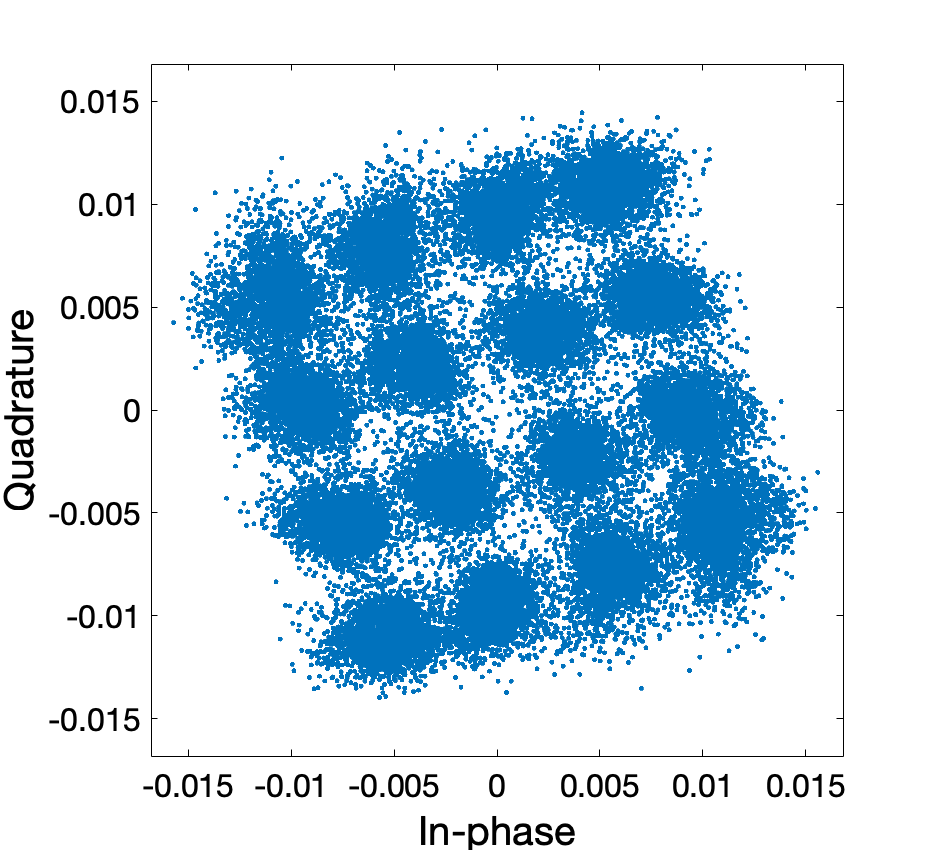}}
    \caption{Constellations for different situations. (The red arrows indicate the distortion in the constellation points, illustrating the effects of expansion, displacement, and rotation caused by hardware imperfections.)}
    \label{fig:constellation}
\end{figure*}

\section{Hardware Imperfection Model}\label{III}
In this section, we introduce each hardware imperfection individually, including imperfect DAC and ADC, IQ imbalance, phase noise, imperfect phase shifters, and nonlinear PA. Then all these hardware imperfections are coupled together to formulate the received signal in the non-ideal MIMO system.
\subsection{Imperfect DAC and ADC Model}
Considering the system with only imperfect DAC and ADC in the Tx and Rx, respectively, we employ the Additive Quantization Noise Model (AQNM) in~\cite{Orhan-2015-Low} to model the imperfection of DAC and ADC, which exhibits good accuracy. Given an input signal $\mathbf{s}$, the output signal $\mathbf{s}_q$ for both ADC and DAC is expressed as~\cite{Orhan-2015-Low}
\begin{equation}
\label{DAC model}
   \mathbf{s}_q=\mathcal{Q}(\mathbf{s})=\alpha\mathbf{s}+\mathbf{q},
\end{equation}
where $\alpha=1-\beta$ is a constant related to the resolution of the DAC and ADC with $\beta\approx \frac{\pi \sqrt {3} }{2}b^{-2}$, and $b$ denotes the number of bits.
Moreover, $\mathbf{q}_{\mathrm{DAC/ADC}}\sim \mathcal{CN}(0,\alpha\beta
\operatorname{diag}(\mathbb{E}(\mathbf{s}\mathbf{s}^\mathrm{H})))$ represents the Gaussian random quantization error~\cite{Orhan-2015-Low} where $\text{diag}(\cdot)$ denotes the diagonal matrix constructed by the diagonal elements of $(\cdot)$. Therefore, the transmitted signal  $\mathbf{x}_{\mathrm{Q}}$ and the received signal $\mathbf{y}_{\mathrm{Q}}$ with imperfect DAC and ADC can be expressed as 
\begin{subequations}
    \begin{align}
    \mathbf{x}_{\mathrm{Q}}&=\mathbf{G}_{\mathrm{PA}}\mathbf{F}_{\mathrm{RF}}\mathbf{P}_{\mathrm{in}}\mathcal{Q}_{\mathrm{DAC}}(\mathbf{F}_{\mathrm{BB}}\mathbf{s})\\
    &=\mathbf{G}_{\mathrm{PA}}\mathbf{F}_{\mathrm{RF}}\mathbf{P}_{\mathrm{in}}(\alpha_{\mathrm{DAC}}\mathbf{F}_{\mathrm{BB}}\mathbf{s}+\mathbf{q}_{\mathrm{DAC}}),
    \end{align}
\end{subequations}
\begin{subequations}
    \begin{align}
    \mathbf{y}_{\mathrm{Q}}&=\mathbf{W}_{\mathrm{BB}}^\mathrm{H} \mathcal{Q}_{\mathrm{ADC}}(\mathbf{W}_{\mathrm{RF}}^\mathrm{H}\mathbf{H}\mathbf{x}_{\mathrm{Q}}+\mathbf{W}_{\mathrm{RF}}^\mathrm{H}\mathbf{n})\\
    &=\mathbf{W}_{\mathrm{BB}}^\mathrm{H} (\alpha_{\mathrm{DAC}}(\mathbf{W}_{\mathrm{RF}}^\mathrm{H}\mathbf{H}\mathbf{x}_{\mathrm{Q}}+\mathbf{W}_{\mathrm{RF}}^\mathrm{H}\mathbf{n})+\mathbf{q}_{\mathrm{ADC}}).
    \end{align}
\end{subequations}
Moreover, $\mathbf{R}_{\mathrm{DAC}}=\alpha_{\mathrm{DAC}}\beta_{\mathrm{DAC}}\text{diag}(\mathbb{E}(\mathbf{F}_{\mathrm{BB}}\mathbf{s}\mathbf{s}^\mathrm{H}\mathbf{F}_{\mathrm{BB}}^\mathrm{H}))$ is the covariance matrix of $\mathbf{q}_{\mathrm{DAC}}$ and the covariance matrix of $\mathbf{q}_{\mathrm{ADC}}$ is $\alpha_{ADC}\beta_{ADC}\text{diag}(\mathbb{E}((\mathbf{W}_{\mathrm{RF}}^H\mathbf{H}\mathbf{x}_{\mathrm{Q}}+\mathbf{W}_{\mathrm{RF}}^H\mathbf{n})(\mathbf{W}_{\mathrm{RF}}^H\mathbf{H}\mathbf{x}_{\mathrm{Q}}+\mathbf{W}_{\mathrm{RF}}^H\mathbf{n})^H))$. 
\subsection{IQ Imbalance Model}
To model the IQ imbalance, we consider the system with only IQ imbalance in the Tx and Rx. The signal $\mathbf{s}$ processed through the digital precoder $\mathbf{F}_{\mathrm{BB}}$ is up-converted through IQ-imbalanced RF chains, which can be given by~\cite{Mahendra-2020-Transmitter}
\begin{equation}
    \label{IQ model}
    \mathcal{Q}_{\mathrm{IQt}}(\mathbf{F}_{\mathrm{BB}}\mathbf{s})=\mathbf{\Gamma}_{\mathrm{t1}}\mathbf{F}_{\mathrm{BB}}\mathbf{s}+\mathbf{\Gamma}_{\mathrm{t2}}(\mathbf{F}_{\mathrm{BB}}\mathbf{s})^*,
\end{equation}
where $\mathbf{\Gamma}_{\mathrm{t1}},\mathbf{\Gamma}_{\mathrm{t2}}\in \mathbb{C}^{L_t\times L_t}$ are the diagonal matrices to model IQ imbalance in the Tx. The $l^{th}$ diagonal elements of $\mathbf{\Gamma}_{\mathrm{t1}},\mathbf{\Gamma}_{\mathrm{t2}}$ are given by
\begin{equation}
        \mathbf{\Gamma}_{\mathrm{t1}}(l,l)=\frac{1+g_{tl}e^{j\phi_{tl}}}{2} \text{ }\text{and}\text{ }
      \mathbf{\Gamma}_{\mathrm{t2}}(l,l)=\frac{1-g_{tl}e^{j\phi_{tl}}}{2},
\end{equation}
where $g_{tl}$ and $\phi_{tl}$ are denoted as the relative amplitude and phase errors between the I and Q branches of $l^{th}$ RF chain in the Tx. Therefore, the transmitted signal $\mathbf{x}_{\mathrm{IQ}}$ with IQ imbalance can be expressed as
\begin{subequations}
    \label{IQ transmitted signal}
    \begin{align}
        \mathbf{x}_{\mathrm{IQ}}&=\mathbf{G}_{\mathrm{PA}}\mathbf{F}_{\mathrm{RF}}\mathbf{P}_{\mathrm{in}}\mathcal{Q}_{\mathrm{IQt}}(\mathbf{F}_{\mathrm{BB}}\mathbf{s})\\
        &=\mathbf{G}_{\mathrm{PA}}\mathbf{F}_{\mathrm{RF}}\mathbf{P}_{\mathrm{in}}(\mathbf{\Gamma}_{\mathrm{t1}}\mathbf{F}_{\mathrm{BB}}\mathbf{s}+\mathbf{\Gamma}_{\mathrm{t2}}(\mathbf{F}_{\mathrm{BB}}\mathbf{s})^*)
    \end{align}
\end{subequations}
Similar to the ideal MIMO system, the digital precoder $\mathbf{F}_{\mathrm{BB}}$ should also satisfy $||\mathbf{F}_{\mathrm{RF}}\mathbf{\Gamma}_{\mathrm{t1}}\mathbf{F}_{\mathrm{BB}}||^2_{\mathrm{F}}+||\mathbf{F}_{\mathrm{RF}}\mathbf{\Gamma}_{\mathrm{t2}}\mathbf{F}_{\mathrm{BB}}^*||^2_{\mathrm{F}}=N_s$ to meet the total transmit power constraint.

At the Rx, the signal processed by the analog combiner is down-converted by the IQ-imbalanced RF chains, which can be modeled as the diagonal matrices $\mathbf{\Gamma}_{\mathrm{r1}},\mathbf{\Gamma}_{\mathrm{r2}}\in \mathbb{C}^{L_r\times L_r}$. The $l_{th}$ diagonal elements of $\mathbf{\Gamma}_{\mathrm{r1}},\mathbf{\Gamma}_{\mathrm{r2}}$ are given by~\cite{Mahendra-2022-Downlink}
\begin{equation}
        \mathbf{\Gamma}_{\mathrm{r1}}(l,l)=\frac{1+g_{rl}e^{j\phi_{rl}}}{2} \text{ }\text{and}\text{ }
      \mathbf{\Gamma}_{\mathrm{r2}}(l,l)=\frac{1-g_{rl}e^{j\phi_{rl}}}{2},
\end{equation}
where $g_{rl}$ and $\phi_{rl}$ are the parameters of I and Q branches of the $l_{th}$ RF chains in the Rx. Then, the received signal $\mathbf{y}_{\mathrm{IQ}}$ with IQ imbalance can be expressed as 
\begin{subequations}
    \begin{align}
        \mathbf{y}_{\mathrm{IQ}}&=\mathbf{W}_{\mathrm{BB}}^\mathrm{H}\mathcal{Q}_{\mathrm{IQr}}(\mathbf{W}_{\mathrm{RF}}^\mathrm{H}\mathbf{H}\mathbf{x}_{\mathrm{IQ}}+\mathbf{W}_{\mathrm{RF}}^\mathrm{H}\mathbf{n})\\
        &=\mathbf{W}_{\mathrm{BB}}^\mathrm{H}(\mathbf{\Gamma}_{\mathrm{r1}}(\mathbf{W}_{\mathrm{RF}}^\mathrm{H}\mathbf{H}\mathbf{x}_{\mathrm{IQ}}+\mathbf{W}_{\mathrm{RF}}^\mathrm{H}\mathbf{n})  \nonumber\\ 
        &+\mathbf{\Gamma}_{\mathrm{r2}}(\mathbf{W}_{\mathrm{RF}}^\mathrm{H}\mathbf{H}\mathbf{x}_{\mathrm{IQ}}+\mathbf{W}_{\mathrm{RF}}^\mathrm{H}\mathbf{n})^*).
    \end{align}
\end{subequations}

 \subsection{Phase Noise Model}
 Considering the imperfect oscillators in the system, the received signal with phase noise can be expressed as~\cite{Forsch-2022-phase}
 \begin{equation}
     \mathbf{y}_\mathrm{PN}=\mathbf{W}_{\mathrm{BB}}^\mathrm{H}\mathbf{\Theta}_{\mathrm{r}}\mathbf{W}_{\mathrm{RF}}^\mathrm{H}\mathbf{H}\mathbf{G}_{\mathrm{PA}}\mathbf{F}_{\mathrm{RF}}\mathbf{\Theta}_{\mathrm{t}}\mathbf{P}_{\mathrm{in}}\mathbf{F}_{\mathrm{BB}}\mathbf{s},
 \end{equation}
 where $\mathbf{\Theta}_{\mathrm{t}}$ $=\text{diag}\{e^{j\theta_1},\cdots,e^{j\theta_{L_t}}\}$ and $\mathbf{\Theta}_{\mathrm{r}}=\text{diag}\{e^{j\theta_1},\cdots,e^{j\theta_{L_r}}\}$ are the phase noise matrices at the Tx and Rx, respectively. 
 The phase noise can be represented as a Wiener process which could be approximated by a Gaussian model for THz bands, i.e., $\theta \sim \mathcal{N}(0,\sigma^2_{\theta})$~\cite{Forsch-2022-phase}. Suppose that each RF chain in the Tx and Rx has its own oscillator, all phase noise terms in  $\mathbf{\Theta}_{\mathrm{t}}$ and  $\mathbf{\Theta}_{\mathrm{r}}$ are independent.
 \subsection{Imperfect Phase Shifter Model}
 To derive the imperfect phase shifter model, we consider the system with only imperfect phase shifters in the Tx and Rx. The input and output signal of the phase shifter can be denoted as $\alpha_{\mathrm{in}}e^{j\phi_{\mathrm{in}}}$ and $\alpha_{\mathrm{out}}e^{j\phi_{\mathrm{out}}}$ respectively. In an ideal case, we would have $\alpha_{\mathrm{in}}=\alpha_{\mathrm{out}}$ and $\phi_{\mathrm{out}}=\phi_{\mathrm{in}}+\Delta_{\phi}$, where $\Delta_{\phi}$ is the phase shift of the b-bit phase shifter belonging to $ [0, 2\pi/2b, \cdots,(2b-1)2\pi/2b]$. However, for an imperfect phase shifter, there could exist amplitude error and phase error~\cite{Kim-2015-220-330}, which can be expressed as
 \begin{align}
     10\log_{10}(\frac{\alpha_{\mathrm{out}}}{\alpha_{\mathrm{in}}})=E_{\alpha},\\
     \phi_{\mathrm{out}}=\phi_{\mathrm{in}}+\Delta_{\phi}+E_{\phi}, 
 \end{align}
 where $E_{\alpha}$ is the amplitude error in dB, and $E_{\phi}$ is the phase error in degree. To account for these errors, we introduce $\mathbf{E}_{\mathrm{F}}$ and $\mathbf{E}_{\mathrm{W}}$ as error matrices in the Tx and Rx, respectively, and each element of $\mathbf{E}_{\mathrm{F}}$ and $\mathbf{E}_{\mathrm{W}}$ can be represented by $10^{\frac{E_\alpha}{10}}e^{jE_{\phi}}$. Therefore, we add the amplitude and phase error and obtain the imperfect analog precoder and combiner, i.e., $\mathbf{F}_{\mathrm{RF,e}}=\mathbf{F}_{\mathrm{RF}}\odot \mathbf{E}_{\mathrm{F}}$ and $\mathbf{W}_{\mathrm{RF}}=\mathbf{W}_{\mathrm{RF}}\odot \mathbf{E}_{\mathrm{W}}$. The transmitted signal $\mathbf{x}_{\mathrm{PS}}$ and received signal $\mathbf{y}_{\mathrm{PS}}$ with imperfect phase shifters can be given by
 \begin{align}
    \mathbf{x}_{\mathrm{PS}}&=\mathbf{G}_{\mathrm{PA}}\mathbf{F}_{\mathrm{RF,e}}\mathbf{P}_{\mathrm{in}}\mathbf{F}_{\mathrm{BB}}\mathbf{s},\\
    \mathbf{y}_{\mathrm{PS}}&=\mathbf{W}_{\mathrm{BB}}^\mathrm{H}\mathbf{W}_{\mathrm{RF,e}}^\mathrm{H}\mathbf{H}\mathbf{x}_{\mathrm{PS}}+\mathbf{W}_{\mathrm{BB}}^\mathrm{H}\mathbf{W}_{\mathrm{RF,e}}^\mathrm{H}\mathbf{n}.
 \end{align}
 \subsection{Nonlinear PA Model}
Considering the system with only nonlinear PAs, we assume that the PAs in the different antennas have the same input-output relation, which is widely used in~\cite{Tervo-2023-Parametrization},~\cite{Moghadam-2018-on} and~\cite{Abdelaziz-2018-digital-predistortion}. We use the Rapp model~\cite{Tervo-2023-Parametrization} to characterize the amplitude-to-amplitude modulation (AM-AM) and amplitude-to-phase modulation (AM-PM) of the nonlinear PAs which are the functions of the power of input signal. We denote $\mathbf{x}_{\mathrm{in}}=\mathbf{F}_{\mathrm{RF}}\mathbf{P}_{\mathrm{in}}\mathbf{F}_{\mathrm{BB}}\mathbf{s}\in\mathbb{C}^{N_t \times T}$ as the input signal of the nonlinear PAs. Then the output signal $\mathbf{x}_{\mathrm{PA}}$ can be expressed as
\begin{align}
     |\mathbf{x}_{\mathrm{PA}}(i,j)|&=\frac{\alpha_{a}}{(1+(\alpha_{a}\frac{|\mathbf{x}_{\mathrm{in}}(i,j)|}{x_{sat}})^{2\sigma_{a}})^{\frac{1}{2\sigma_{a}}}}|\mathbf{x}_{\mathrm{in}}(i,j)|,\\
    \angle{\mathbf{x}_{\mathrm{PA}}(i,j)}&=\frac{\alpha_{\phi}|\mathbf{x}_{\mathrm{in}}(i,j)|^{q_1}}{1+(\frac{|\mathbf{x}_{\mathrm{in}}(i,j)|}{\beta_{\phi}})^{q_2}}+\angle{\mathbf{x}_{\mathrm{in}}(i,j)},\\
    \mathbf{x}_{\mathrm{PA}}(i,j)&=\mathcal{Q}_{\mathrm{PA}}(|\mathbf{x}_{\mathrm{in}}(i,j)|)=|\mathbf{x}_{\mathrm{PA}}(i,j)|e^{j\angle{\mathbf{x}_{\mathrm{PA}}(i,j)}}
\end{align}
where $\alpha_{a}$, $\alpha_{\phi}$, $\beta_{\phi}$, $\sigma_a$, $x_{sat}$, $q_1$ and $q_2$ are model parameters. Therefore, the received signal $\mathbf{y}_{\mathrm{PA}}$ with nonlinear PAs can be expressed as
\begin{equation}
    \mathbf{y}_{\mathrm{PA}}=\mathbf{W}_{\mathrm{BB}}^\mathrm{H} \mathbf{W}_{\mathrm{RF}}^\mathrm{H}\mathbf{H}\mathbf{x}_{\mathrm{PA}}+\mathbf{W}_{\mathrm{BB}}^\mathrm{H} \mathbf{W}_{\mathrm{RF}}^\mathrm{H}\mathbf{n}
\end{equation}

\begin{table*}[ht]
    \centering
    \caption{Comparison on the hardware imperfections}
    \begin{tabular}{c c c}
        \hline
        \textbf{Hardware imperfection}  &\textbf{Characteristics} &\textbf{Impact on constellation} \\
        \hline
        Quantization noise& Random & expansion \\
      
       IQ imbalance& nonlinear &displacement \\
       
        Phase noise &nonlinear & rotation around the origin\\
          
        Amplitude and phase error &Random& expansion \\
       
        PA nonlinearity &nonlinear & complex rotation  \\
        \textbf{All imperfections} &\textbf{random and nonlinear} & \textbf{expansion, rotation and displacement}\\
        \hline
    \end{tabular}
    \label{tab:impact on constellation}
\end{table*}
\subsection{Signal Model with All Imperfect Hardware}
We follow the imperfect transmitted signal and received signal in (\ref{eq:imperfect transmitted signal}) and (\ref{eq:imperfect received signal}) to investigate the signal model with all imperfect hardware step by step. The baseband signal after DACs is denoted as $\mathbf{x}_1$ which can be expressed as
\begin{subequations}
    \begin{align}
    \mathbf{x}_1&=\mathcal{Q}_{\mathrm{DAC}}(\mathbf{F}_{\mathrm{BB}}\mathbf{s})\\&=\alpha_{\mathrm{DAC}}\mathbf{F}_{\mathrm{BB}}\mathbf{s}+\mathbf{q}_{\mathrm{DAC}}
\end{align}
\end{subequations}

Then the signal $\mathbf{x}_1$ is up-converted by the imperfect oscillators with IQ imbalance and phase noise which is represented as
\begin{subequations}
\begin{align}
     \mathbf{x}_2&=\mathbf{\Theta}_{\mathrm{t}}\mathcal{Q}_{\mathrm{IQt}}(\mathbf{P}_{\mathrm{in}}\mathbf{x}_1)\\&=\mathbf{\Theta}_{\mathrm{t}}\mathbf{P}_{\mathrm{in}}(\mathbf{\Gamma}_{\mathrm{t1}}\mathbf{x}_{1}+\mathbf{\Gamma}_{\mathrm{t2}}\mathbf{x}_2^*)
\end{align}
\end{subequations}
where, $\mathbf{x}_2 \in \mathbb{C}^{L_t\times T}$. After up-converted to the THz band, the signal is processed by an analog precoder with imperfect phase shifters and amplified by the nonlinear PAs. Therefore, the non-ideal transmitted signal $\mathbf{x}_e$ can be expressed as
\begin{equation}
    \mathbf{x}_e=\mathcal{Q}_{\mathrm{PA}}(\mathbf{F}_{\mathrm{RF,e}}\mathbf{x}_2)
\end{equation}
After transmitted over the wireless channel $\mathbf{H}$, the signal is processed by the imperfect analog combiner which is given by
\begin{equation}
    \mathbf{y}_1=\mathbf{W}_{\mathrm{RF,e}}^\mathrm{H}\mathbf{H}\mathbf{x}_e+\mathbf{W}_{\mathrm{RF,e}}^\mathrm{H}\mathbf{n}
\end{equation}
Then the received RF signal $\mathbf{y}_1$ is down-converted to baseband signal $\mathbf{y}_2$ with IQ imbalance and phase noise which can be expressed as
\begin{equation}
\label{eq:characterized the imperfect system}
    \mathbf{y}_2=\mathbf{\Theta_{r}}\mathcal{Q}_{\mathrm{IQr}}(\mathbf{y}_1)=\mathbf{\Theta_{r}}(\mathbf{\Gamma}_{\mathrm{r1}}\mathbf{y}_1+\mathbf{\Gamma}_{\mathrm{r2}}\mathbf{y}_1^*)
\end{equation}
Finally, the analog signal $\mathbf{y}_2$ passing through the imperfect ADCs is processed by the digital combiner. Therefore, the imperfect received signal $\mathbf{y}_e$ is
\begin{equation}\label{eq:full hardware_eq}
    \mathbf{y}_e=\mathbf{W}_{\mathrm{BB}}^\mathrm{H}\mathcal{Q}_{\mathrm{ADC}}(\mathbf{y}_2)=\alpha_{\mathrm{ADC}}\mathbf{W}_{\mathrm{BB}}^\mathrm{H}\mathbf{y}_2+\mathbf{W}_{\mathrm{BB}}^\mathrm{H}\mathbf{q}_{\mathrm{ADC}}
\end{equation}

Based on the hardware imperfection models described, we investigate the constellations 
of individual hardware imperfection and their combined imperfections, comparing them to the ideal situation in Fig. \ref{fig:constellation}(a). Fig. \ref{fig:constellation}(b) and (c) illustrate that quantization noise and imperfect phase shifters can cause the expansion of the constellation points. Fig. \ref{fig:constellation}(d) demonstrates that IQ imbalance can cause displacement in different directions. Fig. \ref{fig:constellation}(e) and (f) show that phase noise and nonlinear PAs can induce rotation in the constellation. Specifically, phase noise causes a uniform rotation of the constellation points around the origin whereas PA nonlinearity results in a more complex rotation due to varying degrees of gain and phase distortion. A summary of the impacts of hardware imperfections on the constellation is presented in Table \ref{tab:impact on constellation}.

\begin{figure*} 

    \centering
\includegraphics[scale=0.23]{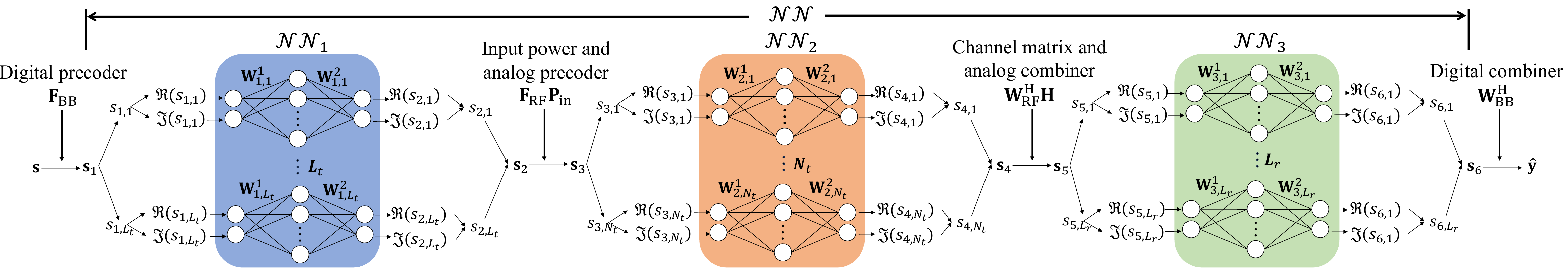}
    \caption{The structure of the proposed DNN in the first stage.}
   \label{fig:hardware modeling}
    \vspace{-5mm}
\end{figure*}

\section{Compensation Problem Formulation}\label{IV}
In this section, we formulate the compensation problem for the hardware imperfections. To address these imperfections, we focus on designing either the baseband digital precoder or combiner to conduct pre- or post-compensation. We denote the digital precoder and combiner for the compensation as $\mathbf{F}_{\mathrm{BB,c}}$ $\in \mathbb{C}^{L_t \times N_s}$ and $\mathbf{W}_{\mathrm{BB,c}}$ $\in \mathbb{C}^{L_r \times N_s}$, respectively. Consequently, the compensated received signals $\mathbf{y}_{ct}$ and $\mathbf{y}_{cr}$ can be expressed as
\begin{equation}
    \label{imperfect received signal}
    \mathbf{y}_{ct}=\mathbf{W}_\mathrm{BB}^\mathrm{H}\mathcal{Q}_{\mathrm{ADC}}(\mathbf{\Theta}_{\mathrm{r}}\mathcal{Q}_{\mathrm{IQr}}(\mathbf{W}_\mathrm{RF,e}^\mathrm{H}\mathbf{H}\mathbf{x}_{ct}+\mathbf{W}_\mathrm{RF,e}^\mathrm{H}\mathbf{n})),
\end{equation}    
\begin{equation}
    \label{imperfect received signal}
    \mathbf{y}_{cr}=\mathbf{W}_\mathrm{BB,c}^\mathrm{H}\mathcal{Q}_{\mathrm{ADC}}(\mathbf{\Theta}_{\mathrm{r}}\mathcal{Q}_{\mathrm{IQr}}(\mathbf{W}_\mathrm{RF,e}^\mathrm{H}\mathbf{H}\mathbf{x}_e+\mathbf{W}_\mathrm{RF,e}^\mathrm{H}\mathbf{n})),
\end{equation}   
where
\begin{equation}
    \label{imperfect transmitted signal}
    \mathbf{x}_{ct}=\mathcal{Q}_{\mathrm{PA}}(\mathbf{F}_{\mathrm{RF,e}}\mathbf{\Theta}_{\mathrm{t}}\mathcal{Q}_{\mathrm{IQt}}( \mathbf{P}_{\mathrm{in}}\mathcal{Q}_{\mathrm{DAC}}(\mathbf{F}_\mathrm{BB,c}\mathbf{s}))).
\end{equation}
Then, we can obtain the transmitted signal $\mathbf{s}_{ct(cr)}$, which can be given as
\begin{equation}
   \mathbf{s}_{ct(cr)}=(\mathbf{W}_{\mathrm{BB}}^\mathrm{H} \mathbf{W}_{\mathrm{RF}}^\mathrm{H}\mathbf{H}\mathbf{F}_{\mathrm{RF}}\mathbf{F}_{\mathrm{BB}})^{-1}\mathbf{y}_{ct(cr)}.
\end{equation}
The aim of the compensation problem is to design either a new digital precoder or combiner to minimize the mean square error (MSE) between  $\mathbf{s}_{ct(cr)}$ and $\mathbf{s}$, which can be formulated as 
\begin{subequations}
    \begin{align}
         &\min_{\mathbf{F}_{\mathrm{BB,c}}\  \text{or} \ \mathbf{W}_{\mathrm{BB,c}}} \mathrm{MSE}\\
\text{s.t.}&||\mathbf{F}_{\mathrm{RF}}\mathbf{F}_{\mathrm{BB,c}}||^2_{\mathrm{F}}=N_s\\&||\mathbf{W}_{\mathrm{BB,c}}^\mathrm{H}\mathbf{W}_{\mathrm{RF}}^\mathrm{H}||^2_{\mathrm{F}}=N_s
    \end{align}
\end{subequations}
where
\begin{equation}
  \mathrm{MSE}=\frac{1}{MN}\sum_{m=1}^{M}\sum_{n=1}^{N}(\mathbf{s}_{ct(cr)}(m,n)-\mathbf{s}(m,n))^2
\end{equation}

\section{Two-stage Hardware Imperfection Compensation Algorithm With DNN}\label{V}
In Sec. \ref{III}, we provide mathematical solutions for all hardware imperfections. Although an approximate analytical solution can be derived from the mathematical solutions to compensate for the imperfections, it is challenging to obtain precise information and model parameters for all hardware imperfections in practical communication systems. Therefore, in this section, we propose a two-stage hardware imperfection compensation algorithm using deep learning to solve the formulated compensation problem. In the first stage, we develop a DNN, which is composed of multiple sub-neural networks (sub-NNs) to represent the hardware imperfections based on the signal transmission process. In the second stage, utilizing the DNN developed in the first stage, we design either the digital precoder at the Tx or the combiner at the Rx using NN to effectively compensate for the hardware imperfections.
\subsection{Stage~1: DNN for Hardware Imperfection Representation}
As shown in Fig.~\ref{fig:hardware modeling}, the DNN consists of three parts, i.e., $\mathcal{NN}_1$, $\mathcal{NN}_2$ and $\mathcal{NN}_3$ to represent the combined hardware imperfections in the Tx RF chains, Tx antennas, and Rx RF chains, respectively. Each part is composed of many identical sub-NNs. Each sub-NN has two input nodes for the real and imaginary parts of the input symbol, a single hidden layer with $N_h$ hidden nodes, and two output nodes for the real and imaginary parts of the output symbol.
We note that the DNN is used to represent the system characterized by (\ref{eq:characterized the imperfect system}). Therefore, we design the NN based on the signal transmission process, i.e., using $L_t$ sub-NNs in $\mathcal{NN}_1$, $N_t$ sub-NNs in $\mathcal{NN}_2$ and $L_r$ sub-NNs in $\mathcal{NN}_3$ to represent the imperfections in $L_t$ Tx RF chains, $N_t$ transmitted antennas and $L_r$ Rx RF chains, respectively.

In Fig.~\ref{fig:hardware modeling}, The input of $\mathcal{NN}_1$ is the baseband signal $\mathbf{s}_1=\mathbf{F}_{\mathrm{BB}}\mathbf{s} \in \mathbb{C}^{L_t\times T}$. Specifically, $\mathbf{s}_1$ can be decomposed as $\mathbf{s}_1=[\mathbf{s}_{1,1}^\mathrm{T},\cdots,\mathbf{s}_{1,l_t}^\mathrm{T},\cdots,\mathbf{s}_{1,L_t}^\mathrm{T}]^\mathrm{T}$ where $\mathbf{s}_{1,l_t} \in \mathbb{C}^{1\times T}$. The real part $\mathrm{Re}(\mathbf{s}_{1,l_t})\in\mathbb{N}^{1\times T}$ and the imaginary part $\mathrm{Im}(\mathbf{s}_{1,l_t})\in\mathbb{N}^{1\times T}$ are input to the $l_t^{th}$ sub-NN in $\mathcal{NN}_1$. Moreover, the activation function \textit{Tanh} is adopted in a hidden layer which determines the activated neurons in the network. In particular, the \textit{Tanh} activation function is represented as
\begin{equation}
    f_\mathrm{Tanh}(x)=\frac{1-\mathrm{exp}(-2x)}{1+\mathrm{exp}(-2x)}
\end{equation}
Therefore, the input to the $l_t^{th}$ sub-NN in $\mathcal{NN}_1$ is denoted as
\begin{equation}
    \mathbf{c}_{1,l_t}=[\mathrm{Re}({s}_{1,l_t}),\mathrm{Im}({s}_{1,l_t})]^\mathrm{T}.
\end{equation}
The output of the $l_r^{th}$ sub-NN hide layer is
\begin{equation}
    \mathbf{h}_{1,l_t}=f_\mathrm{Tanh}(\mathbf{W}^1_{1,l_t}\mathbf{c}_{1,l_t}+\mathbf{b}^1_{1,l_t}),
\end{equation}
where $\mathbf{W}^1_{1,l_t}\in \mathbb{C}^{N_h\times 2}$ and $\mathbf{b}^1_{1,l_t}\in \mathbb{C}^{N_h\times 1}$ are the corresponding weight matrix and bias vector of the $l_r^{th}$ sub-NN hide layer, and $\mathbf{W}^1_{1,l_t}=[\mathbf{w}^1_{1,l_t,1},\mathbf{w}^1_{1,l_t,2}]^\mathrm{T}$ with $\mathbf{w}^1_{1,l_t,1}=[{w}^1_{1,l_t,11},\cdots,{w}^1_{1,l_t,N_h1}]^\mathrm{T}$ and $\mathbf{w}^1_{1,l_t,2}=[{w}^1_{1,l_t,12},\cdots,{w}^1_{1,l_t,N_h2}]^\mathrm{T}$. The output of the $l_r^{th}$ sub-NN in $\mathcal{NN}_1$ is expressed as
\begin{equation}
s_{2,l_t}=(\mathbf{W}^{2}_{1,l_t})^\mathrm{T}\mathbf{h}_{1,l_t}+\mathbf{b}^2_{1,l_t},
\end{equation}
where $\mathbf{W}^2_{1,l_t}\in \mathbb{C}^{N_h\times2}$ and $\mathbf{b}^2_{1,l_t}\in \mathbb{C}^{N_h\times 1}$ are the corresponding weight matrix and bias vector of the $l_t^{th}$ sub-NN output layer. Therefore, we can obtain $\mathbf{s}_2=[s_{2,1},\cdots,s_{2,l_t},\cdots,s_{2,L_t}]^\mathrm{T}\in \mathbb{C}^{L_t\times1}$ which is the output of $\mathcal{NN}_1$. For simplicity, we denote $\mathcal{NN}_1$ as a function, i.e., $\mathbf{s}_2=\mathcal{NN}_1(\mathbf{s}_1)$.

The input of $\mathcal{NN}_2$ to represent the imperfections in transmitted antennas can be expressed as
\begin{equation}
    \mathbf{s}_3=\mathbf{F}_{\mathrm{RF}}\mathbf{P}_{\mathrm{in}}\mathbf{s}_2,
\end{equation}
where $\mathbf{s}_3\in \mathbb{C}^{N_t \times 1}$ can be decomposed as $\mathbf{s}_3=[s_{3,1},\cdots,s_{3,n_t},\cdots,s_{3,N_t}]^\mathrm{T}$. To characterize the nonlinearity, the \textit{Tanh} is used in the hide layers of sub-NNs in $\mathcal{NN}_2$. Therefore, the input to the $n_t^{th}$ sub-NN in $\mathcal{NN}_2$ can be denoted as
\begin{equation}
    \mathbf{c}_{2,l_t}=[\mathrm{Re}({s}_{3,n_t}),\mathrm{Im}({s}_{3,n_t})]^\mathrm{T}
\end{equation}
The hidden layer output in $n_t^{th}$ sub-NN is
\begin{equation}
     \mathbf{h}_{2,n_t}=f_\mathrm{Tanh}(\mathbf{W}^1_{2,n_t}\mathbf{c}_{2,n_t}+\mathbf{b}^1_{2,n_t}),
\end{equation}
where $\mathbf{W}^1_{2,n_t}\in \mathbb{C}^{N_h\times 2}$ and $\mathbf{b}^1_{2,n_t}\in \mathbb{C}^{N_h\times 1}$ are the corresponding weight matrix and bias vector of the $n_t^{th}$ sub-NN hide layer, and $\mathbf{W}^1_{2,n_t}=[\mathbf{w}^1_{2,n_t,1},\mathbf{w}^1_{2,n_t,2}]^\mathrm{T}$ with $\mathbf{w}^1_{2,n_t,1}=[{w}^1_{2,n_t,11},\cdots,{w}^1_{2,n_t,N_h1}]^\mathrm{T}$ and $\mathbf{w}^1_{2,n_t,2}=[{w}^1_{2,n_t,12},\cdots,{w}^1_{2,n_t,N_h2}]^\mathrm{T}$. The output of the $n_t^{th}$ sub-NN in $\mathcal{NN}_2$ is expressed as
\begin{equation}
s_{4,n_t}=(\mathbf{W}^{2}_{2,n_t})^\mathrm{T}\mathbf{h}_{2,n_t}+\mathbf{b}^2_{2,n_t},
\end{equation}
where $\mathbf{W}^2_{2,n_t}\in \mathbb{C}^{N_h\times2}$ and $\mathbf{b}^2_{2,n_t}\in \mathbb{C}^{N_h\times 1}$ are the corresponding weight matrix and bias vector of the $n_t^{th}$ sub-NN output layer. Therefore, we can obtain $\mathbf{s}_4=[s_{4,1},\cdots,s_{4,n_t},\cdots,s_{4,N_t}]^\mathrm{T}\in \mathbb{C}^{N_t\times1}$ which is the output of $\mathcal{NN}_2$ and can be expressed as $\mathbf{s}_4=\mathcal{NN}_2(\mathbf{s}_3)$.

Based on the signal transmission process, $\mathbf{s}_4$ passes through the channel $\mathbf{H}$ and analog combiner $\mathbf{W}_{\mathrm{RF}}$ and $\mathbf{s}_5=\mathbf{W}_{\mathrm{RF}}^\mathrm{H} \mathbf{H} \mathbf{s}_4$ can be obtained which is the input of $\mathcal{NN}_3$. Similar to the process in $\mathcal{NN}_1$, the input and output for the $L_r^{th}$ sub-NNs is 
\begin{equation}
      \mathbf{c}_{3,l_t}=[\mathrm{Re}({s}_{5,l_t}),\mathrm{Im}({s}_{5,l_t})]^\mathrm{T},
\end{equation}
\begin{equation}
s_{6,l_r}=(\mathbf{W}^{2}_{3,l_r})^\mathrm{T}\mathbf{h}_{3,l_r}+\mathbf{b}^2_{3,l_r},
\end{equation}
where
\begin{equation}
    \mathbf{h}_{3,l_r}=f_\mathrm{Tanh}(\mathbf{W}^1_{3,l_r}\mathbf{c}_{3,l_r}+\mathbf{b}^1_{3,l_r}).
\end{equation}
$\mathbf{W}^1_{3,l_r}\in \mathbb{C}^{N_h\times2}$ and $\mathbf{b}^1_{3,l_r}\in \mathbb{C}^{N_h\times 1}$ are the corresponding weight matrix and bias vector of the $l_r^{th}$ sub-NN hidden layer. $\mathbf{W}^2_{3,l_r}\in \mathbb{C}^{N_h\times2}$ and $\mathbf{b}^2_{3,l_r}\in \mathbb{C}^{N_h\times 1}$ are the corresponding weight matrix and bias vector of the $l_r^{th}$ sub-NN output layer. Then, the process can be described as $\mathbf{s}_6=\mathcal{NN}_3(\mathbf{s}_5)$. Finally, the represented signal can be expressed as $\hat{\mathbf{y}}_e=\mathbf{W}_{\mathrm{BB}}^\mathrm{H}\mathbf{s}_6$.

In DNN, the training process is straightforward. We denote the length of pilot signal as $M$, i.e., we have the $M$ training pairs $\{(\mathbf{y}_e(m),\mathbf{s}(m)),m=1,\cdots,M\}$. 
During the training process, the MSE loss function denoted by $L_\mathrm{MSE}$ is deployed, which is defined as 
\begin{equation}
    \mathrm{L_{MSE}}=\frac{1}{L_r}\frac{1}{M}\sum_{l_r=1}^{L_r}\sum_{m=1}^{M}(\hat{\mathbf{y}}_e(l_r,m)-\mathbf{y}_e(l_r,m))^2,
\end{equation}
The weights $\{\mathbf{W}^{1}_{1,l_r}, \mathbf{W}^{2}_{1,l_r},\mathbf{W}^{1}_{2,n_t},\mathbf{W}^{2}_{2,n_t},\mathbf{W}^{1}_{3,l_r},\mathbf{W}^{2}_{3,l_r}\}$ and biases$\{\mathbf{b}^{1}_{1,l_r}, \mathbf{b}^{2}_{1,l_r},\mathbf{b}^{1}_{2,n_t},\mathbf{b}^{2}_{2,n_t},\mathbf{b}^{1}_{3,l_r},\mathbf{b}^{2}_{3,l_r}\}$ are determined in the back-propagation process. 

After training, we can obtain the hardware imperfection representation which can be expressed as
\begin{align}
    \hat{\mathbf{y}}
&=\mathbf{W}_{\mathrm{BB}}^\mathrm{H}\mathcal{NN}_3(\mathbf{W}_{\mathrm{RF}}^\mathrm{H}\mathbf{H}\mathcal{NN}_2(\mathbf{F}_{\mathrm{RF}}\mathbf{P}_{\mathrm{in}}\mathcal{NN}_1(\mathbf{s}_1)))\\&=\mathbf{W}_{\mathrm{BB}}^\mathrm{H}\mathcal{NN}(\mathbf{F}_{\mathrm{BB}}\mathbf{s}),
\end{align}
where $\mathcal{NN}$ is denoted as the trained DNN.

\begin{figure}[t]
    \centering
    \subfigure[NN for the Tx compensation.]{\includegraphics[width=3.2in]{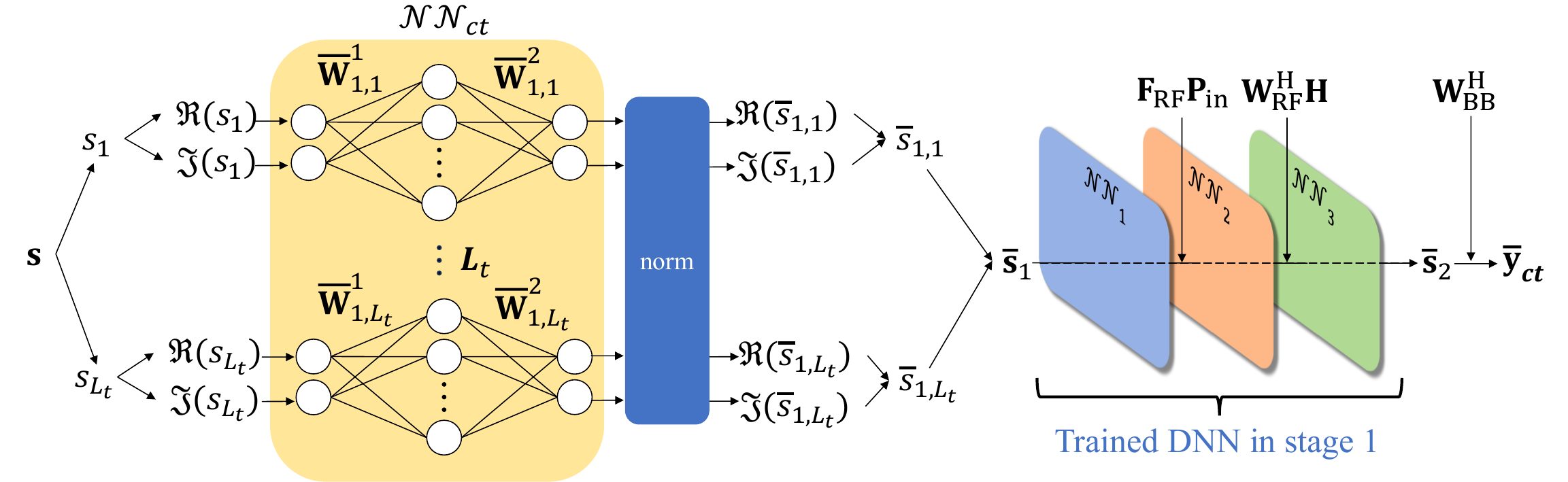}}
    \subfigure[NN for the Rx compensation.]{\includegraphics[width=3.2in]{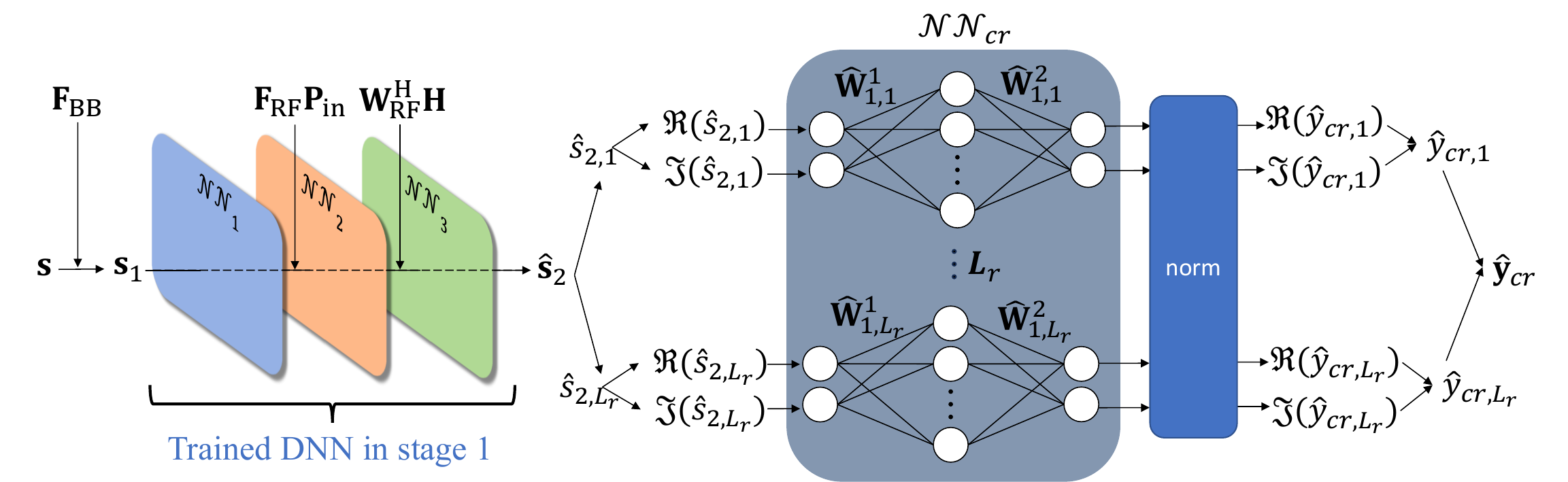}} \\
    \caption{The structure of the proposed NN in the second stage.}
      \label{fig:baseband compensation}
\end{figure}
\subsection{Stage~2: Compensation Scheme with NN}
Since we obtain the hardware imperfection model $\mathcal{NN}(\cdot)$, the baseband compensation problem can be reformulated as
\begin{subequations}
\begin{align}
    \min_{\mathbf{F}_{\mathrm{BB,c}}\  \text{or} \ \mathbf{W}_{\mathrm{BB,c}}} &\frac{1}{MN_s}\sum_{i=1}^{N_s}\sum_{j=1}^{M}(\hat{\mathbf{y}}_{ct(cr)}(i,j)-\mathbf{y}_i(i,j))^2\\
\text{s.t.}&||\mathbf{F}_{\mathrm{RF}}\mathbf{F}_{\mathrm{BB,c}}||^2_{\mathrm{F}}=N_s\\&||\mathbf{W}_{\mathrm{BB,c}}^\mathrm{H}\mathbf{W}_{\mathrm{RF}}^\mathrm{H}||^2_{\mathrm{F}}=N_s
\end{align}
\end{subequations}
where
\begin{equation}
    \hat{\mathbf{y}}_{ct}=\mathbf{W}_{\mathrm{BB}}^\mathrm{H}\mathcal{NN}(\mathbf{F}_{\mathrm{BB,c}}\mathbf{s})
\end{equation}
and
\begin{equation}
    \hat{\mathbf{y}}_{cr}=\mathbf{W}_{\mathrm{BB,c}}^\mathrm{H}\mathcal{NN}(\mathbf{F}_{\mathrm{BB}}\mathbf{s}).
\end{equation}
As shown in Fig. \ref{fig:baseband compensation}(a), we depict the process of designing $\mathbf{F}_{\mathrm{BB,c}}$ or $\mathbf{W}_{\mathrm{BB,c}}$ to compensate for hardware imperfections. In the Tx, the digital precoded signal $\mathbf{s}_1$ is regarded as the input of the transmitted baseband design $\mathcal{NN}_{ct}$ which consists of $L_t$ sub-NNs. Therefore, the input and output for the $L_t^{th}$ sub-NNs in $\mathcal{NN}_{ct}$ is
\begin{equation}
      \overline{\mathbf{c}}_{1,l_t}=[\mathrm{Re}({s}_{l_t}),\mathrm{Im}({s}_{l_t})]^\mathrm{T},
\end{equation}
\begin{equation}
 \overline{s}_{1,l_t}= (\overline{\mathbf{W}}^{2}_{1,l_t})^\mathrm{T} \overline{\mathbf{h}}_{1,l_t}+ \overline{\mathbf{b}}^2_{1,l_t},
\end{equation}
where 
\begin{equation}
     \overline{\mathbf{h}}_{1,l_t}=f_\mathrm{Tanh}( \overline{\mathbf{W}}^1_{1,l_t} \overline{\mathbf{c}}_{1,l_t}+ \overline{\mathbf{b}}^1_{1,l_t}).
\end{equation}
$\overline{\mathbf{W}}^1_{1,l_t}\in \mathbb{C}^{N_h\times2}$ and $\overline{\mathbf{b}}^1_{1,l_t}\in \mathbb{C}^{N_h\times 1}$ are the corresponding weight matrix and bias vector of the $l_r^{th}$ sub-NN hidden layer. $\overline{\mathbf{W}}^2_{1,l_t}\in \mathbb{C}^{N_h\times2}$ and $\overline{\mathbf{b}}^2_{1,l_t}\in \mathbb{C}^{N_h\times 1}$ are the corresponding weight matrix and bias vector of the $l_r^{th}$ sub-NN output layer. Due to the power constraint in the baseband, the output $\overline{\mathbf{s}}_1$ should be normalized to satisfied $||\overline{\mathbf{s}}_1||_{\mathrm{F}}^2=||\mathbf{s}||_{\mathrm{F}}^2=N_s$. Then $\overline{\mathbf{s}}_1$ is input to the trained DNN $\mathcal{NN}(\cdot)$ and the output is $\overline{\mathbf{s}}_2=\mathcal{NN}(\overline{\mathbf{s}}_1)$. Finally, the signal compensated by the digital precoder can be expressed as $\overline{\mathbf{y}}_{ct}=\mathbf{W}_{\mathrm{BB}}^\mathrm{H}\overline{\mathbf{s}}_2$.

In $\mathcal{NN}_{ct}$, the training process is straightforward. With $M$ pilot signals, we have the $M$ training pairs $\{(\mathbf{y}_i(m),\mathbf{s}(m)), m=1,\cdots, M\}$ where $\mathbf{y}_i$ is the ideal signal calculate in \eqref{eq:ideal received signal}. During the training process, the MSE loss function is deployed, which is defined as
\begin{equation}
     \mathrm{L_{MSE}}^{ct}=\frac{1}{L_r}\frac{1}{M}\sum_{l_r=1}^{L_r}\sum_{m=1}^{M}(\overline{\mathbf{y}}_{ct}(l_r,m)-\mathbf{y}_{i}(l_r,m))^2,
\end{equation}
The weights $\{\overline{\mathbf{W}}^{1}_{1,l_t}, \overline{\mathbf{W}}^{2}_{1,l_t}\}$ and biases$\{\overline{\mathbf{b}}^{1}_{1,l_t}, \overline{\mathbf{b}}^{2}_{1,l_t}\}$ are determined with back-propagation. 
Since $\overline{\mathbf{s}}_{1}=\mathbf{F}_{\mathrm{BB,c}}\mathbf{s}$, the digital precoder for hardware compensation can be obtained by $\mathbf{F}_{\mathrm{BB,c}}=\overline{\mathbf{s}}_{1}\mathbf{s}^{-1}$.

As shown in Fig. \ref{fig:baseband compensation}(b), to compensate in the Rx, the input of the received baseband design network $\mathcal{NN}_{cr}$ is $\hat{\mathbf{s}}_2=\mathcal{NN}(\mathbf{s}_1)$. With $L_r$ sub-NNs in $\mathcal{NN}_{cr}$, the input and output for the $L_r^{th}$ sub-NNs in $\mathcal{NN}_{cr}$ is
\begin{equation}
      \hat{\mathbf{c}}_{1,l_t}=[\mathrm{Re}(\hat{\mathbf{s}}_{2,l_r}),\mathrm{Im}(\hat{\mathbf{s}}_{2,l_r})]^\mathrm{T},
\end{equation}
\begin{equation}
 \hat{\mathbf{y}}_{cr,l_r}= (\hat{\mathbf{W}}^{2}_{1,l_r})^\mathrm{T} \hat{\mathbf{h}}_{1,l_r}+ \hat{\mathbf{b}}^2_{1,l_r},
\end{equation}
where
\begin{equation}
     \hat{\mathbf{h}}_{1,l_r}=f_\mathrm{Tanh}( \hat{\mathbf{W}}^1_{1,l_r} \hat{\mathbf{c}}_{1,l_r}+ \hat{\mathbf{b}}^1_{1,l_r}).
\end{equation}
With the power constraint in the baseband, the output $\hat{\mathbf{y}}_{cr}$ should be normalized to satisfied $||\hat{\mathbf{y}}_{cr}||_{\mathrm{F}}^2=||\hat{\mathbf{s}}_2||_{\mathrm{F}}^2=N_s$ and the digital combiner is $\mathbf{W}_{\mathrm{BB}}=(\hat{\mathbf{y}}_{cr}\hat{\mathbf{s}}_2^{-1})^\mathrm{H}$. Similar to the training process in the Tx compensation, we have the $M$ training pairs $\{(\mathbf{y}_i(m),\hat{\mathbf{s}}_2(m)), m=1,\cdots, M\}$ and the MSE loss function is
\begin{equation}
     \mathrm{L_{MSE}}^{cr}=\frac{1}{L_r}\frac{1}{M}\sum_{l_r=1}^{L_r}\sum_{m=1}^{M}(\hat{\mathbf{y}}_{cr}(l_r,m)-\mathbf{y}_{i}(l_r,m))^2,
\end{equation}

\section{Network slimming for DNN in the first phase}\label{VI}
Since the hardware imperfection compensation is based on the trained DNN in the first stage, it is a prerequisite for effective compensation to obtain an accurate extraction of hardware imperfection characteristics. By carefully designing the neural network, we aim to accomplish tasks with the fewest possible parameters, thereby reducing computational complexity while maintaining good performance. Therefore, we propose three methods to slim the DNN in the first stage.
\begin{figure}[ht]
    \centering
\includegraphics[scale=0.29]{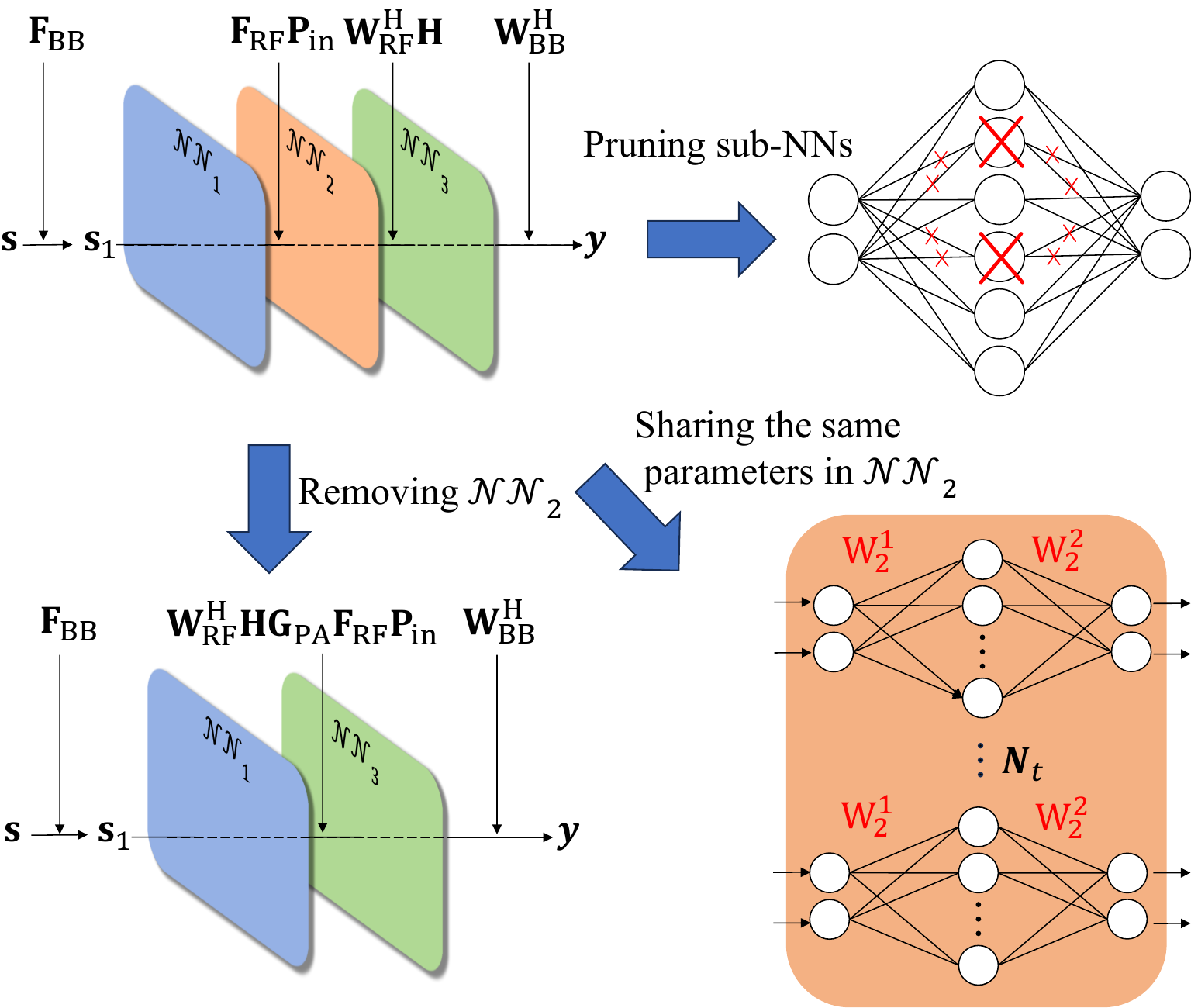}
    \caption{Three methods of network slimming.}
    \label{fig:slim}
    \vspace{-5mm}
\end{figure}

\subsection{Pruning}
Firstly, we calculate the number of the parameters in the sub-NN. From the input layer to the hidden layer, the numbers of the weight and bias parameters are $2N_h$ and $N_h$, respectively. From the hidden layer to the output layer, the numbers of the weight and bias parameters are $2N_h$ and $2$. Therefore, for a sub-NN with $2$ input neurons, $N_h$ hidden neurons, and $2$ output neurons, the total number of the parameters in a sub-NN is $2N_h+N_h+2N_h+2=5N_h+2$. The numbers of the sub-NNs in $\mathcal{NN}_1$, $\mathcal{NN}_2$, $\mathcal{NN}_3$ are $L_t$, $N_t$ and $L_r$, respectively. Then the total number of the parameters in $\mathcal{NN}$ is $(L_t+N_t+L_r)(5N_h+2)$. Although increasing the number of hidden neurons can improve the modeling of the hardware information, it also increases the training time, computational resources, and the risk of overfitting. 

Hence, it is essential to perform appropriate pruning on the sub-NNs by reducing the number of neurons in the hidden layers, as shown in Fig. \ref{fig:slim}.
On the one hand, by reducing the number of hidden layer neurons in the sub-NNs, the computational load during training is reduced, making the neural network more efficient and can be used on resource-constrained devices. On the other hand, pruning the sub-NNs can address the risk of overfitting that arises from overly complex models with excessive parameters. Pruning the sub-NNs encourages the network to focus on the most significant features and extract the features from data affected by various types of Gaussian noise.

\subsection{Parameter Sharing}
In the hardware imperfection representation network $\mathcal{NN}$, $\mathcal{NN}_2$ plays a role in characterizing the hardware imperfection of the nonlinear PAs in the transmit antenna array. Since all nonlinear PAs in BS have the same hardware feature, all sub-NNs in $\mathcal{NN}_2$ can share the same weight and bias parameters shown in Fig. \ref{fig:slim}. This parameter sharing strategy significantly reduces the overall complexity and training time of $\mathcal{NN}$, shrinking the total number of parameters from $(L_t+N_t+L_r)(5N_h+2)$ to $(L_t+1+L_r)(5N_h+2)$. Parameter sharing in $\mathcal{NN}_2$ not only reduces the computational complexity but also enhances the training efficiency. By leveraging shared parameters, $\mathcal{NN}_2$ can capture the common patterns of PA nonlinearity, promoting robust performance with different datasets and avoiding the overfitting problem. 

\begin{figure}[t]
    \centering
\includegraphics[width=2.8in]{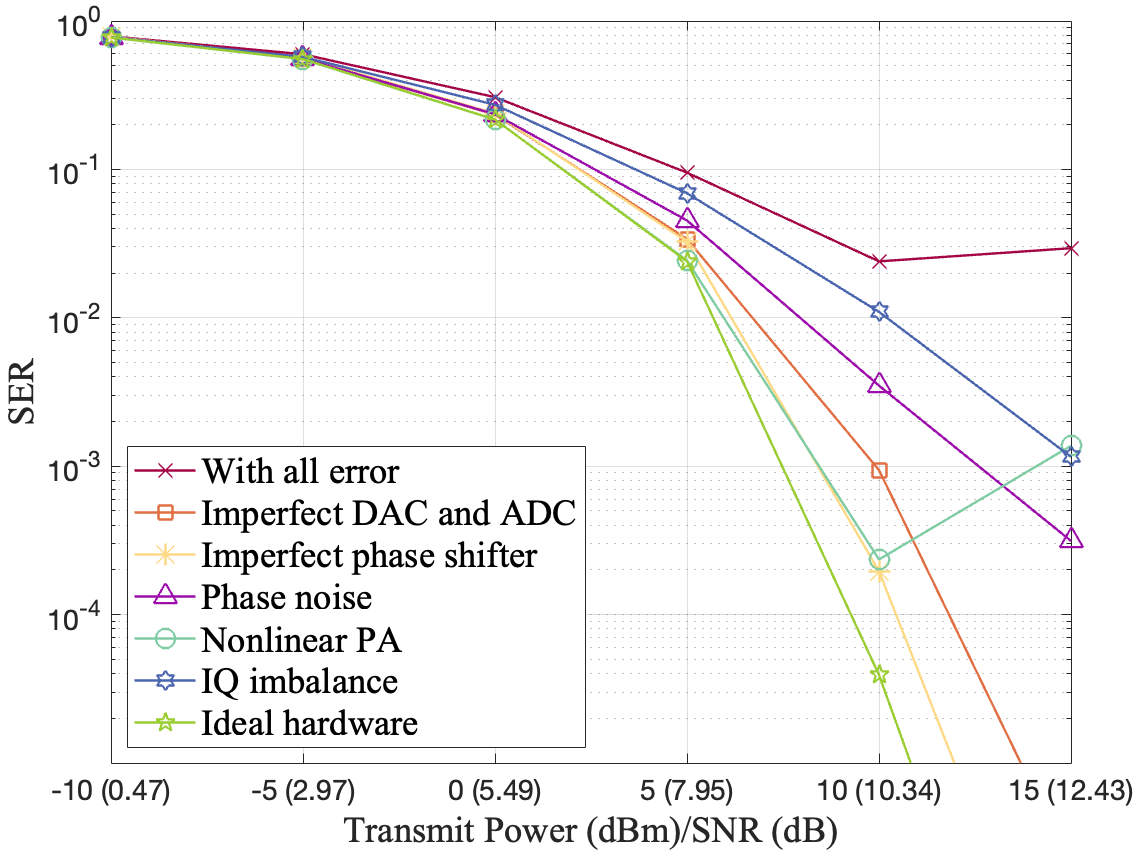}
    \caption{The SER performance of the system with ideal hardware, each imperfect hardware, and all imperfect hardware.}
    \label{fig:SER}
    \vspace{-5mm}
\end{figure}

\subsection{Removing $\mathcal{NN}_2$ at Low Transmit Power}

When transmit power is low, the nonlinearity of PAs is reduced, simplifying the system hardware imperfections and making them easier to capture. Therefore, employing a relatively simple neural network is sufficient to represent these imperfections, significantly reducing computational complexity while mitigating the risk of overfitting the problem.
As illustrated in Fig. \ref{fig:slim}, since $\mathcal{NN}_2$ is designed to characterize PA nonlinearity, it can be omitted under low transmit power conditions, as the nonlinearity of the PA is not significant at low transmit power. Then, the total number of parameters is reduced from $(L_t+N_t+L_r)(5N_h+2)$ to $(L_t+L_r)(5N_h+2)$. The model for hardware imperfections can be expressed as:
\begin{equation}
\mathbf{y} = \mathbf{W}_{\mathrm{BB}}^\mathrm{H} \mathcal{NN}_3 \left( \mathbf{W}_{\mathrm{RF}}^\mathrm{H} \mathbf{H} \mathbf{G}_\mathrm{PA}\mathbf{F}_{\mathrm{RF}}  \mathbf{P}_{\mathrm{in}} \mathcal{NN}_1(\mathbf{s}_1) \right),
\end{equation}
where $\mathbf{G}_\mathrm{PA}$ represents the PA gain at low input power. When transmit power levels increase, $\mathcal{NN}_2$ can be reintroduced into $\mathcal{NN}$ to ensure effective modeling of hardware imperfections, demonstrating the flexibility of $\mathcal{NN}_2$.

\subsection{Combination of three slimming methods}
Since each of the aforementioned three slimming methods can individually reduce the computational complexity of DNN while maintaining good performance, we can also integrate them to achieve further slimming. First, we can decrease the number of hidden layer neurons in all sub-NNs to prune DNN. Next, we implement the parameter sharing method in $\mathcal{NN}_2$ to further decrease the number of parameters. Finally, when the transmit power is reduced to a level that does not cause nonlinearity in the PAs, we can remove $\mathcal{NN}_2$ to simplify DNN.

\begin{figure*}[t]
    \centering
   
    \subfigure[Training loss of stage~1 and stage~2.]{\includegraphics[width=2.4in]{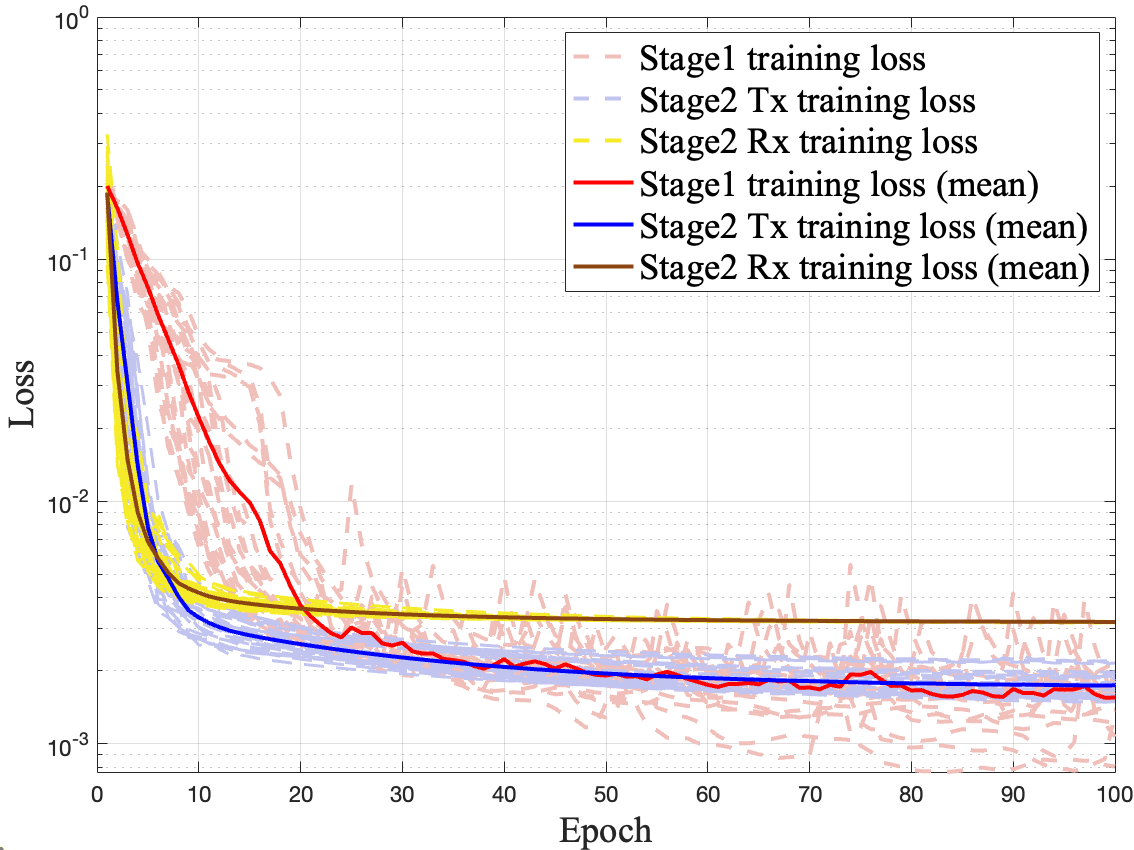}} 
     \subfigure[SER of the Tx and Rx compensation.]{\includegraphics[width=2.3in]{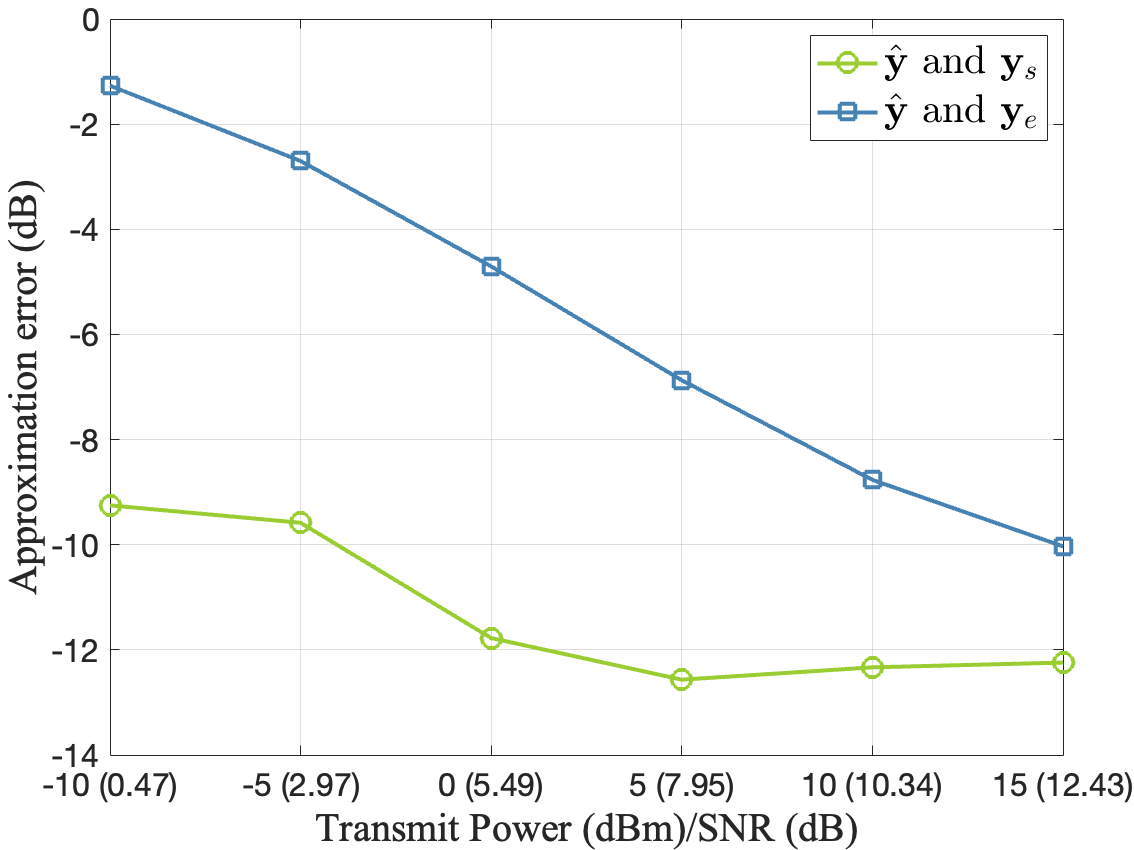}}
     \subfigure[SER of the Tx and Rx compensation.]{\includegraphics[width=2.3in]{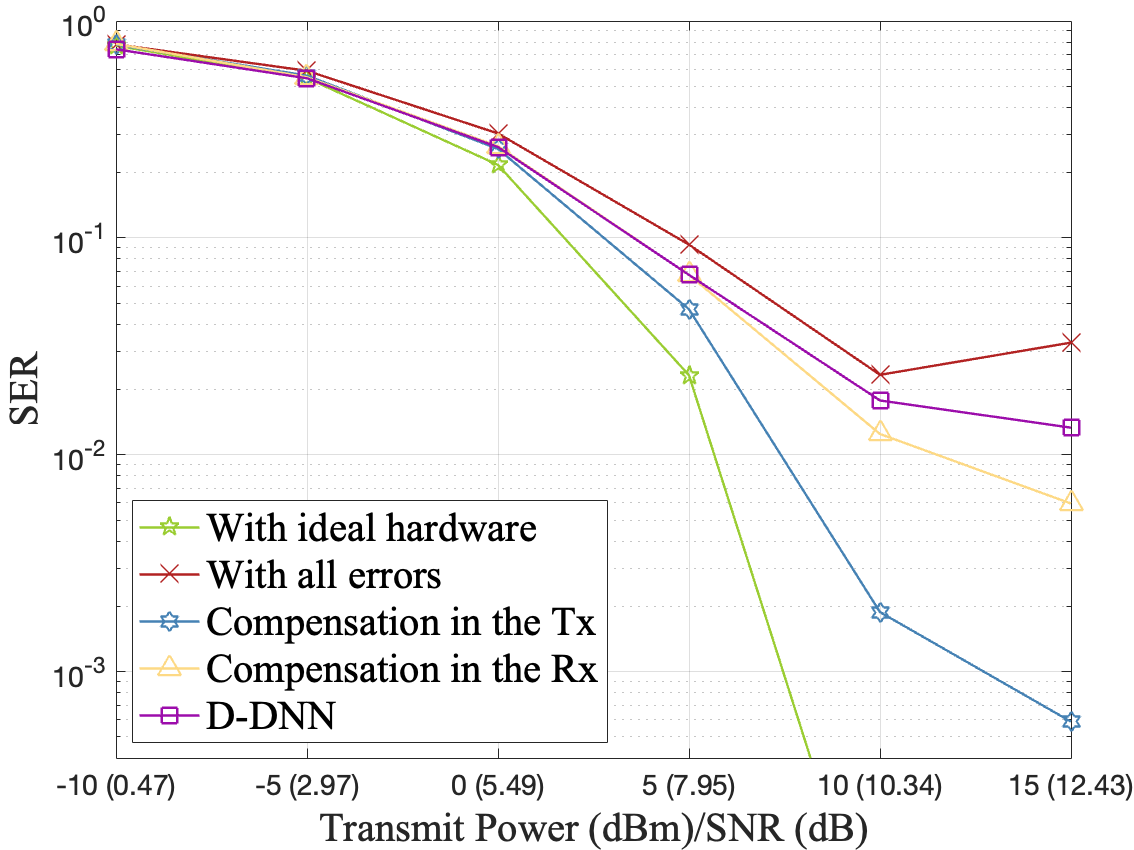}}
    \caption{The performance of hardware imperfection compensation algorithm.}
    \label{fig:performance}
    
\end{figure*}

\begin{figure}[ht]
    \centering
    \subfigure[Imperfect hardware.]{\includegraphics[width=1.5in]{images/constellation/real_con.png}}
    \subfigure[DNN representation.]{\includegraphics[width=1.5in]{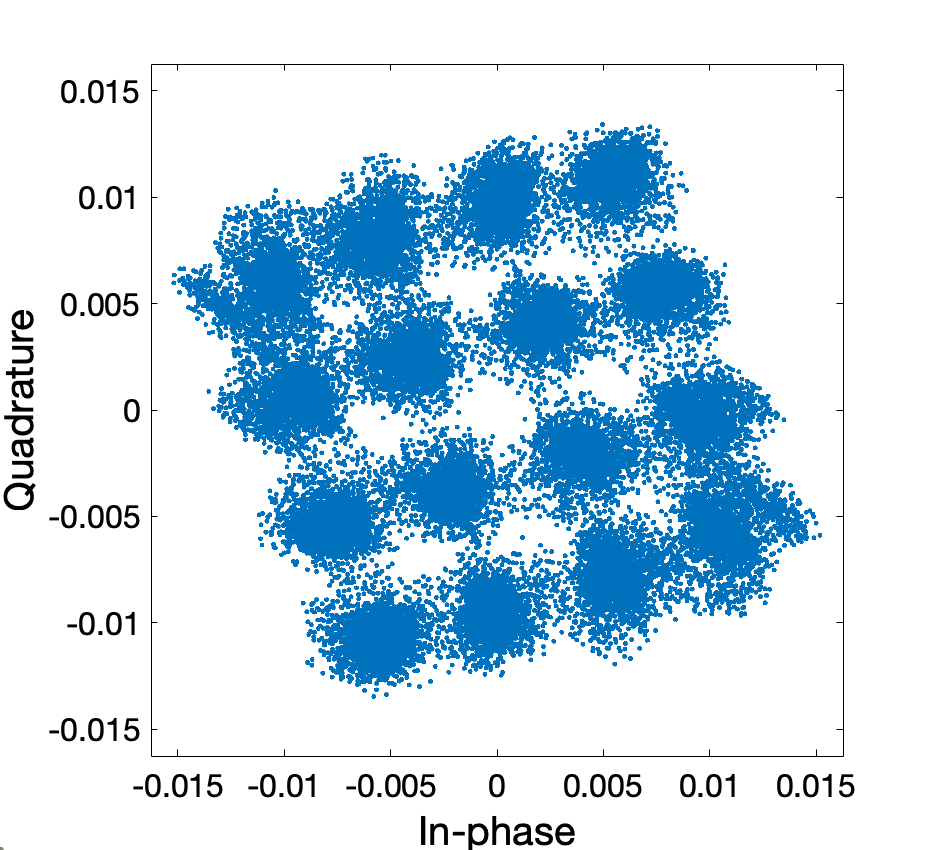}}\\
    \subfigure[Tx compensation.]{\includegraphics[width=1.5in]{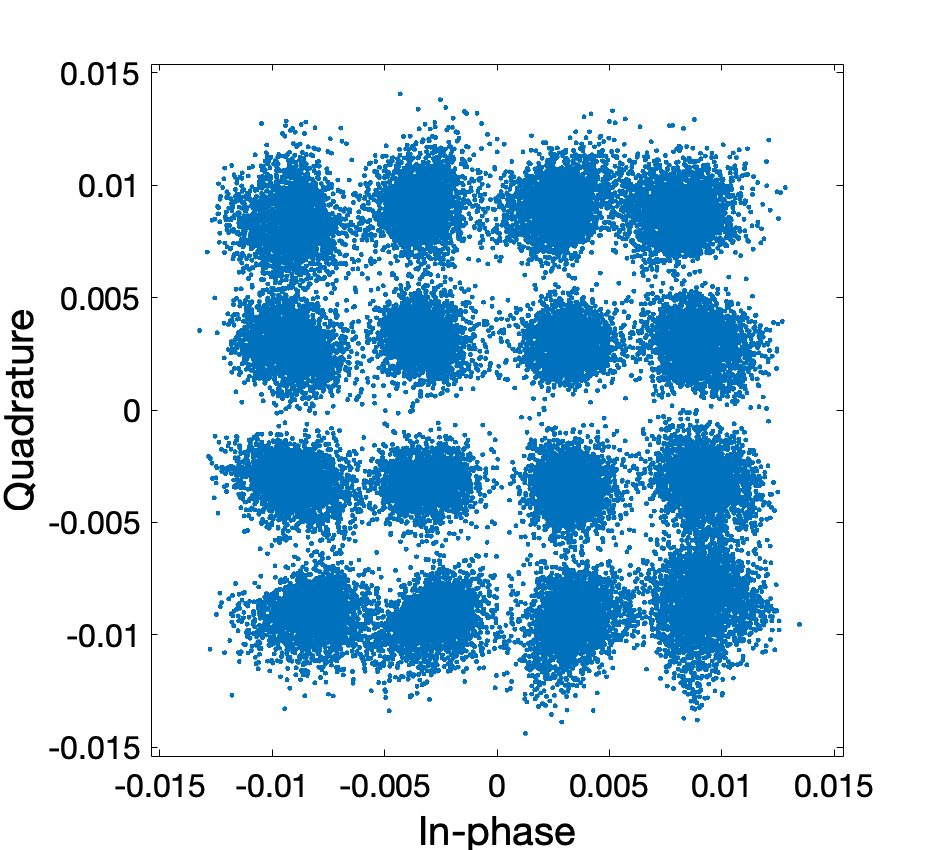}} 
    \subfigure[Rx compensation.]{\includegraphics[width=1.5in]{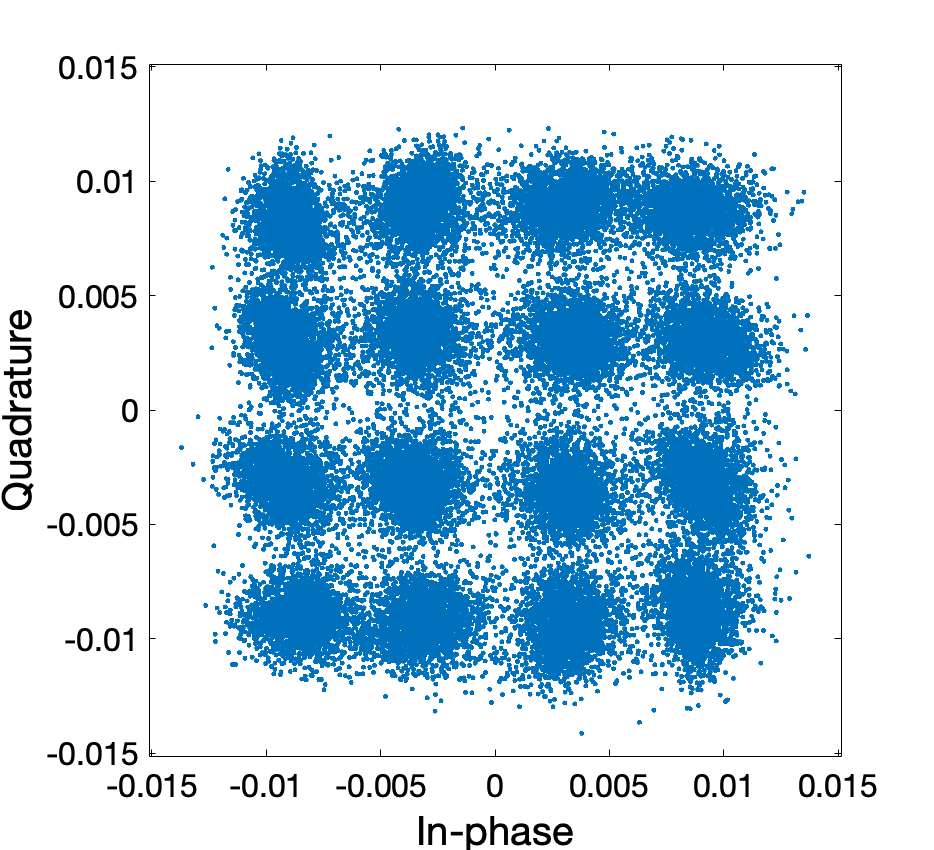}}\\
   
    \caption{The constellation of hardware imperfection compensation algorithm.}
     \label{fig:4constellation}
\end{figure}

\section{Performance Evaluation}\label{VII}
In this section, we first introduce the simulation parameters. Next, the hardware imperfections are analyzed in terms of SER. Furthermore, we evaluate the performance of the compensation algorithm and network slimming methods in both uncoded and coded systems.



\begin{figure*}[ht]
    \centering
    \subfigure[Pruning.]{\includegraphics[width=2.3in]{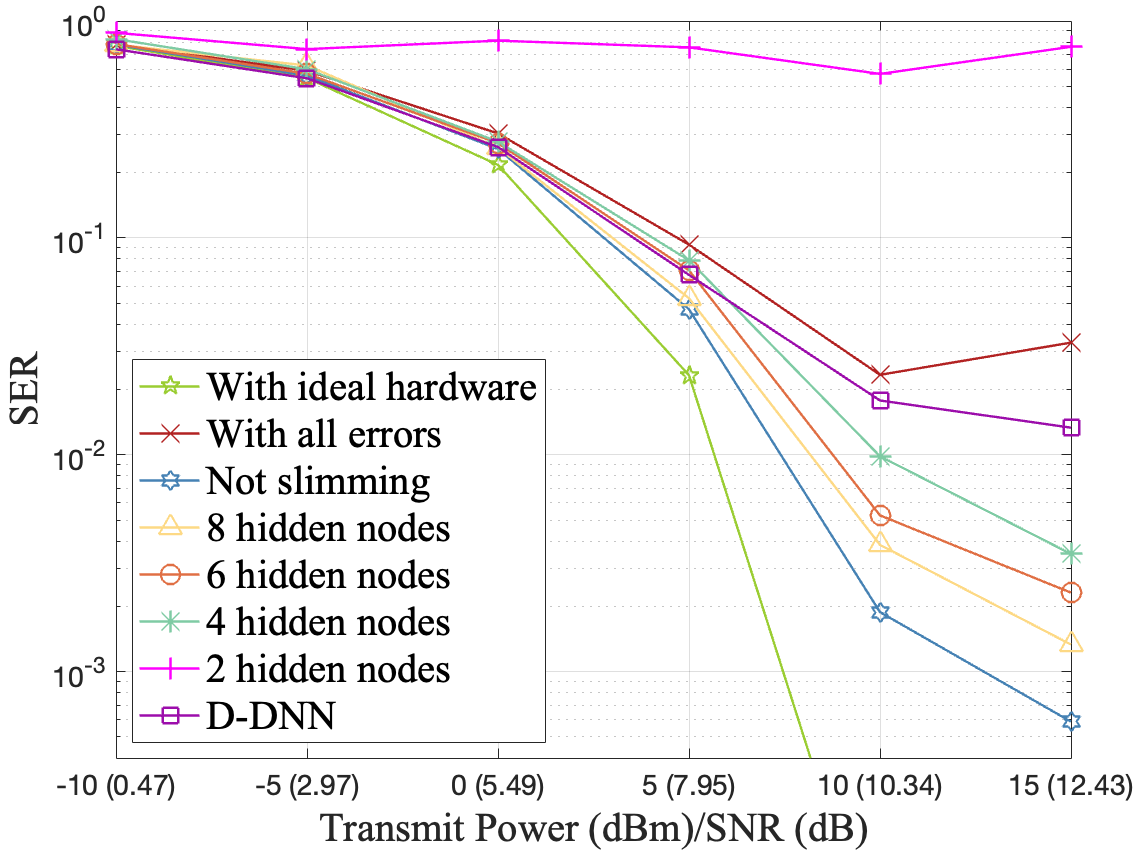}}
    \subfigure[Parameter sharing and removing.]{\includegraphics[width=2.3in]{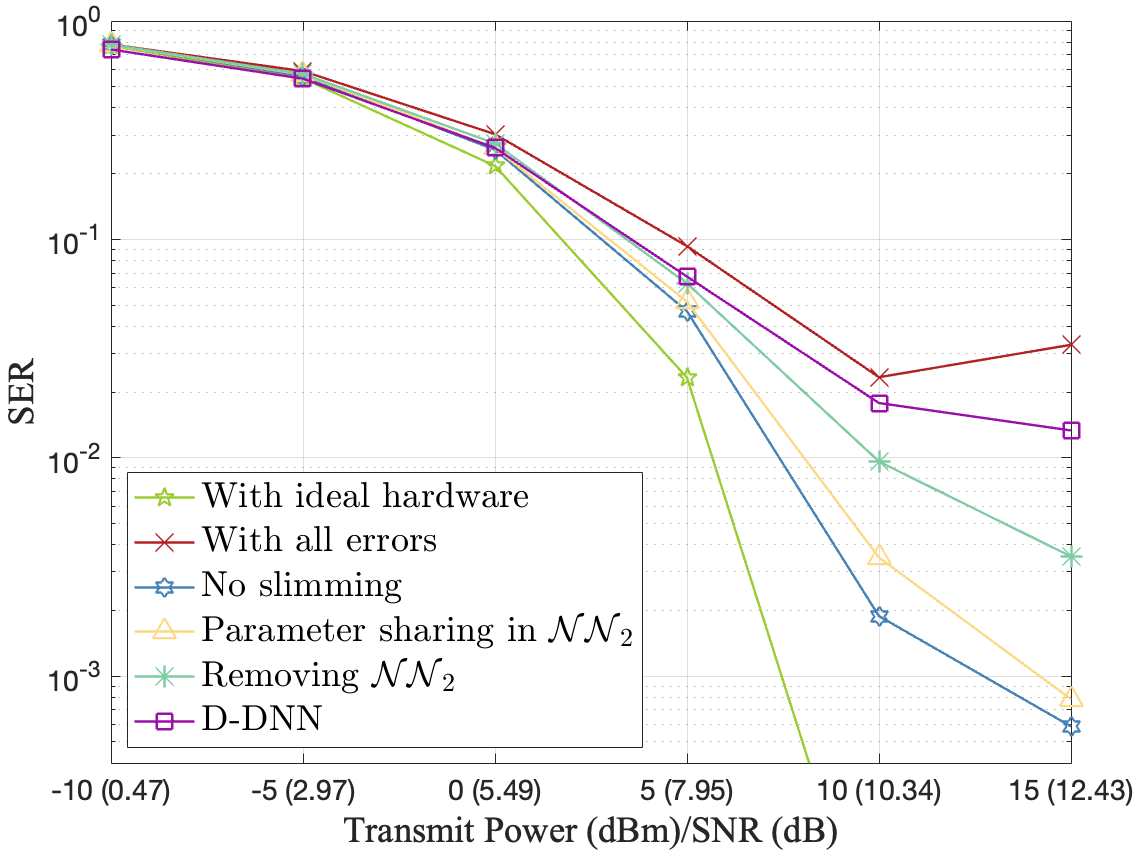}}
     \subfigure[Combination of the slimming methods.]{\includegraphics[width=2.3in]{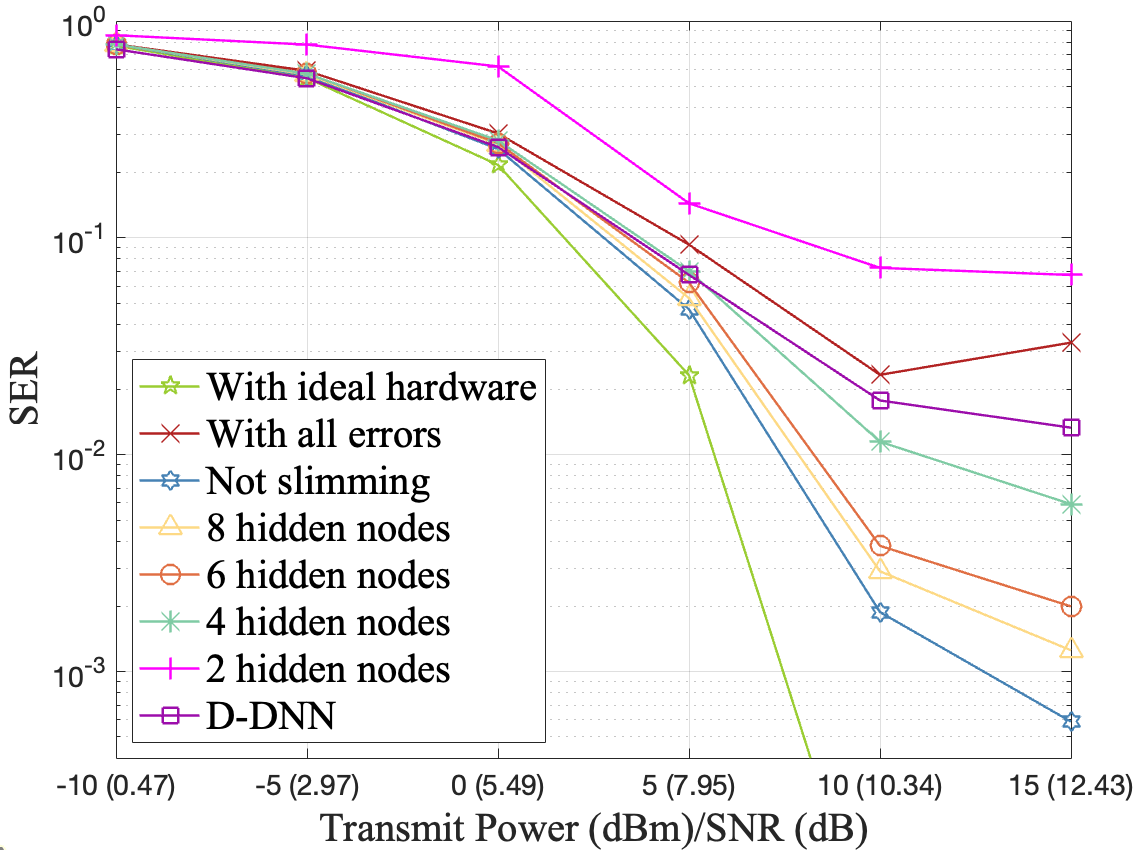}}
    \caption{The performance of hardware imperfection compensation algorithm with different slimming methods.}
     \label{fig:slimming performance}
\end{figure*}


  
  

\subsection{Evaluation Setup}

\textit{1) Simulation parameters:}
The simulation parameters can be divided into two parts: system parameters and hardware imperfection parameters. For system parameters, the numbers of antennas and RF chains at the Tx and Rx are $N_t=N_r=256$ and $L_t=L_r=4$, respectively. Moreover, the number of the data stream is $N_s=4$ and the transmit power ranges from -10 dBm to 15 dBm. The bandwidth is set as $B=$ 1 GHz and the distance between the Tx and Rx is 50m. To provide detailed information about SNR with different transmit power in the figure, we calculate it as $\mathrm{SNR}=||\mathbf{H}\mathbf{x}||_{\mathrm{F}}^2/{B\sigma^2}$,
where $\mathbf{H}$, $\mathbf{x}$, and $\sigma^2$ are channel matrix, transmit signal and noise variance, respectively.
For the hardware imperfections, the resolution of the DAC and ADC is 4 bits. The parameter of the IQ imbalance at the Tx and Rx are $g_{tl}$, $g_{rl}$ $\sim \textit{U}$ [0.9, 1,1], and $\phi_{tl}$, $\phi_{rl}$ $\sim \textit{U}$ $[-20^\circ, 20^\circ]$~\cite{Mahendra-2020-Transmitter}. The variance of the phase noise at the Tx and Rx is $\sigma_\phi^2=10^{-2}$~\cite{Forsch-2022-phase}. The resolution of the phase shifter at the Tx and Rx is 6 bits with amplitude error $E_\alpha \sim \mathcal{N}(0,1.2)$ and phase error $E_\phi \sim \mathcal{N}(0,10.2^\circ)$~\cite{Kim-2015-220-330}. The parameters used for describing PA nonlinearity are $\alpha_a=4.708$, $x_{sat}=0.663$, $\sigma_a=1.603$, $\alpha_\phi=-740.2$, $q_1=1.945$, $\beta_\phi=0.298$ and $q_2=1.797$~\cite{Tervo-2023-Parametrization} and the ideal linear gain of PA is 13 dB.

\textit{2) Training and Testing:}
As it is challenging to obtain data from a real THz hybrid beamforming system with hardware imperfections, we generate the training data through simulation. Specifically, we use a 16-QAM modulation scheme to obtain transmit symbols. According to \eqref{eq:full hardware_eq}, the transmit symbols pass through the hardware imperfection process to obtain the 8000 train data for the first stage. 
In this dataset, the parameters defining the nonlinearities in IQ imbalance and PAs are selected initially to establish a fixed nonlinear function, which remains constant throughout the dataset generation process. Similarly, the amplitude and phase errors in the phase shifters are randomly generated initially and held fixed, as these are attributed to manufacturing imperfections. In contrast, quantization noise and phase noise vary across different symbols, reflecting their inherent randomness in practical systems.
Then according to \eqref{eq:ideal received signal}, we generate 8000 ideal received signals as the train data for the second stage based on the known input power, digital precoder, analog precoder, digital combiner, analog combiner, and channel state information. To form the test data set, we select another 8000 transmit symbols with transmit power ranging from $-10$~dBm to 15~dBm. 
In addition, since the characteristics of the hardware imperfection are different with different transmit power, the neural network needs to be retrained at different transmit power. Moreover, it is worth noting that, although we are currently using simulation-generated data from Sec.   ~\ref{III}, in practice, training the DNN to represent the combined hardware imperfections only requires the pilot information, the corresponding signals received at the receiver, and other prior knowledge, without the need to know the models of individual hardware imperfections.

All the deep learning experimental results are implemented on a PC with Intel(R) Xeon(R) CPU E5-2690 v4 @ 2.60 GHz and an Nvidia GeForce RTX 2080 Ti GPU. In addition, the simulation of the UM-MIMO system and the generation of the training and testing data are operated in Matlab (R2023a) environment, while the proposed DNN and the slimming method are implemented in the Visual Studio framework.
\subsection{Hardware Imperfection Analysis}
The SER performance of the system with ideal hardware, each imperfect hardware, and all imperfect hardware are presented in Fig. \ref{fig:SER}. With the increase of the transmit power, the SER curves in different cases decrease except for the nonlinear PA case, where the power ranging from 10~dBm to 15~dBm strengthens PA nonlinearity, causing its SER to increase. When the transmit power ranges from -10 dBm to -5 dBm, all the curves almost have the same value, which means that the Gaussian noise in the receiver dominates over the hardware imperfection. When the transmit power ranges from 0 dBm to 10 dBm, the SER curves for the phase noise and IQ imbalance are higher compared to those for imperfect phase shifter, nonlinear PA, DAC, and ADC. This indicates that the phase noise and IQ imbalance dominate in this case. When the transmit power ranges from 10 dBm to 15 dBm, nonlinear PA becomes dominant as the input power approaches the saturation power. This leads to a sudden increase in SER between 10~dBm to 15~dBm.

\subsection{Performance of Compensation Algorithms}
\textit{1) Convergence Evaluation:}
The convergence performance of the proposed hardware imperfection compensation method is evaluated in Fig. \ref{fig:performance}(a) by analyzing the convergence speeds of both stage~1 and stage~2. We perform multiple times network training and record the corresponding training loss in the first 100 training epochs. For stage~1, the loss reaches 0.002 at 20 epochs and stabilizes after 40 epochs. For stage~2, the losses for compensation at both the Tx and Rx converge after 20 epochs while the mean loss value at the Tx is 0.0019, which is smaller than the mean loss value of 0.003 at the Rx.

\textit{2) Representation Accuracy:}
To verify the DNN in stage~1 can represent the coupled imperfections from the received signal under the random noise, we compare the signals $\hat{\mathbf{y}}$ output from DNN in stage~1 with the actual received signal $\mathbf{y}_e$ and the signal $\mathbf{y}_s$, where the effect of random noise has been removed. The approximation error of these two cases can be expressed as $||\hat{\mathbf{y}}-\mathbf{y}_e||_{\mathrm{F}}/||\mathbf{y}_e||_{\mathrm{F}}$ and $||\hat{\mathbf{y}}-\mathbf{y}_s||_{\mathrm{F}}/||\mathbf{y}_s||_{\mathrm{F}}$. As shown in Fig. \ref{fig:performance}(b), with the decreasing of the transmit power, the approximation error with $\mathbf{y}_e$ is increasing due to the decrease in SNR. However, the approximation error with $\mathbf{y}_s$ is lower than the approximation error with $\mathbf{y}_e$, which means that DNN in stage~1 can represent the combined imperfections from the signals affected by various types of random noise.

\textit{3) Compensation Performance:}
 We evaluate the compensation performance in terms of the SER and constellation. In Fig. \ref{fig:performance}(c), we compare the compensation applied at the Tx and Rx. When the transmit power is less than 5~dBm, the performance of compensation in the Tx and Rx is similar. However, when the transmit power exceeds 5~dBm, the performance of compensation in the Tx significantly outperforms that in the Rx. Specifically, the compensation in the Tx can achieve a SER of 0.0018 and 0.0005 with 10~dBm and 15~dBm transmit power, while the compensation in the Rx can only achieve a SER of 0.012 and 0.006 with 10~dBm and 15~dBm transmit power. This is because, when compensation is performed at the transmitter, the parameters input to the compensation neural network come from a fixed set of signals modulated using 16QAM, which are consistent during both training and testing of the neural network. However, when compensation is performed at the receiver, the parameters input to the compensation neural network are influenced by varying random noise, and there are differences between the training signals and the actual received signals. Therefore, it is preferable to perform hardware imperfection compensation in the Tx. Furthermore, we also compared the proposed methods with the direct-DNN (D-DNN) approach \cite{Jaraut-2018-Composite}, a commonly used method for addressing complex coupling problems. The D-DNN method takes the received pilot signals as input, processes them through a fully connected neural network, and outputs the corrected pilot signals. We evaluate the performance of the D-DNN by setting the input layer to 8 neurons, 3 hidden layers with 10 neurons each, and an output layer of 8 neurons. As illustrated in Fig. \ref{fig:performance}(c), the D-DNN has a similar performance with the proposed methods when the transmit power is less than 0~dBm. However, when the transmit power exceeds 5~dBm, the performance of the D-DNN is notably inferior to that of the proposed algorithms, particularly at 15 dBm transmit power, where it achieves a SER of 0.0133. These results demonstrate the high capability of the proposed compensation to deal with the combined hardware imperfections.
 
 As shown in Fig.~\ref{fig:4constellation}, we present the constellations with 15~dBm transmit power. Fig.~\ref{fig:4constellation}(a) shows the constellation with all imperfect hardware and Fig.~\ref{fig:4constellation}(b) is the constellation obtained from DNN in stage~1. Then Fig.~\ref{fig:4constellation}(c) and (d) present the constellations after the compensation in the Tx and Rx, which indicates that the compensation in the Tx performs better than that in the Rx.

\subsection{Network Slimming}
In Fig.~\ref{fig:slimming performance}, we compare the performance across various slimming methods.
Additionally, Table~\ref{tab:number of parameter} presents the corresponding number of parameters for each method. 
In Fig.~\ref{fig:slimming performance}(a), compared with no slimming network with $N_h=10$, we can find that at 15~dBm transmit power, the SER increases by 0.0007, 0.0017 and 0.0029 when the number of parameters degrades by 19.2\%, 38.2\% and 57.5\% with 8, 6 and 4 hidden layer neurons, respectively.
When the number of hidden layer neurons is reduced to 2, the number of parameters decreases by 76.9\%, making the neural network too simple to capture hardware imperfections, which leads to poor performance in terms of SER.

Moreover, we analyze the other two slimming methods. 
As shown in Fig.~\ref{fig:slimming performance}(b), at 15~dBm, sharing the parameters in $\mathcal{NN}_2$ results in a minimal SER increase of 0.0002. In contrast, removing $\mathcal{NN}_2$ leads to a more significant SER increase of 0.003. However, when the transmit power is below 5~dBm, the performance of the two slimming methods and the no-slimming method is comparable with the removal of $\mathcal{NN}_2$ yielding the simplest network. Thus, when the transmit power is lower than 5 dBm, we can remove $\mathcal{NN}_2$ to reduce the network complexity while maintaining good performance. 

Then, we evaluate the combination of the slimming methods. When the transmit power is not less than 5~dBm, we apply pruning and parameter sharing in $\mathcal{NN}_2$. When the transmit power is lower than 5~dBm, we conduct pruning and removing $\mathcal{NN}_2$.
Fig.~\ref{fig:slimming performance}(c) demonstrates that at 15~dBm transmit power, compared to the no-slimming method, the combined method of parameter sharing and pruning results in an SER increase of only 0.00067, 0.00142, and 0.00532 when the number of parameters is reduced by 97.2\%, 97.9\%, and 98.6\% with 8, 6, and 4 hidden layer neurons, respectively. At 5~dBm transmit power, compared to the no-slimming method, the combined method of removing $\mathcal{NN}_2$ and pruning results in an SER increase of only 0.006, 0.014, and 0.0232 when the number of parameters is reduced by 97.5\%, 98.1\%, and 98.7\% with 8, 6, and 4 hidden layer neurons, respectively. Furthermore, the combined slimming methods with no fewer than 4 hidden nodes all outperform the method using D-DNN.


\subsection{Coded System}
\begin{figure}[t]
    \centering
\includegraphics[width=2.5in]{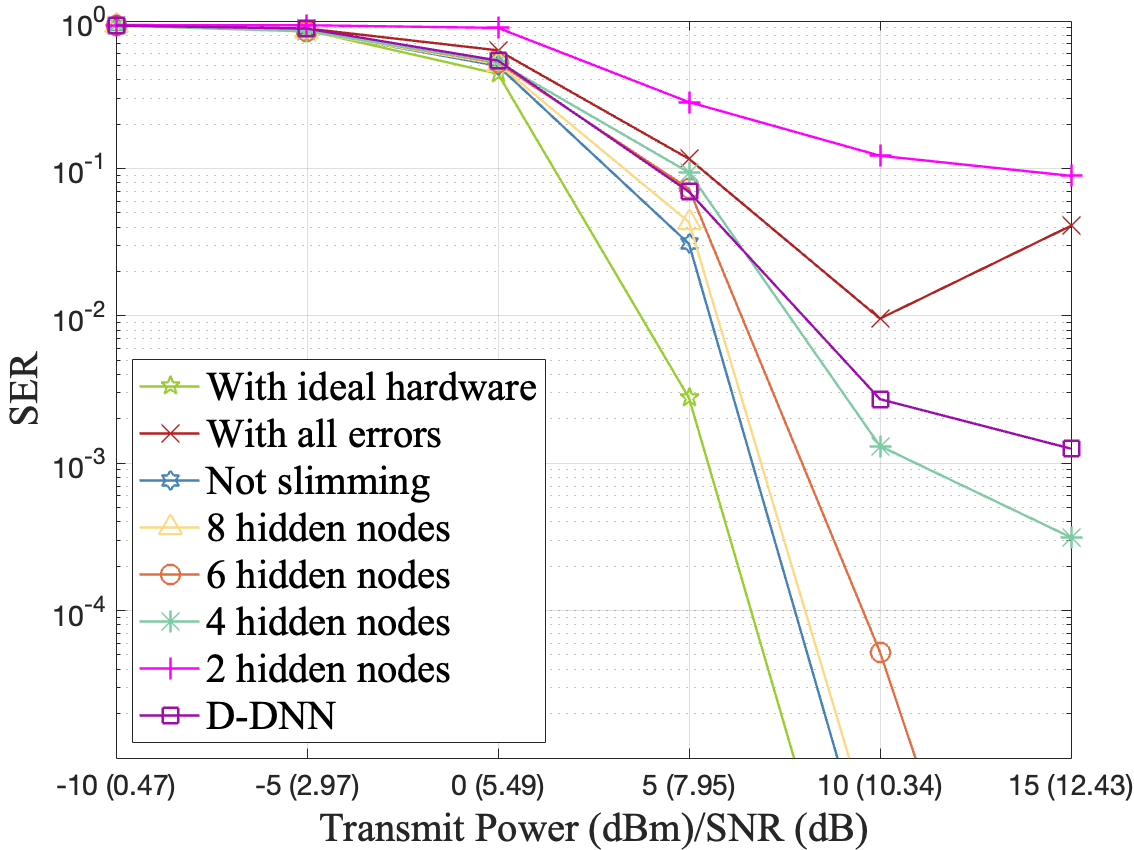}
    \caption{The SER performance of the coded system with combined slimming methods.}
    \label{fig:SER_coded}
    \vspace{-5mm}
\end{figure}

In the previous evaluation, we assessed each method using an uncoded system. Then we evaluate the performance of the proposed methods in a coded system to examine the effectiveness of the proposed methods under realistic coding and modulation conditions. We employ a convolutional code with a code rate of 2/3 and utilize 16QAM modulation with Gray mapping. At the receiver, the outputs after hard decision are fed into a Viterbi decoder to recover the transmitted symbols. We compare the performance of the ideal system, non-ideal system, non-ideal system with combined slimming methods, and non-ideal system with D-DNN. As shown in Fig. \ref{fig:SER_coded}, We can see that the combined slimming methods with no fewer than 6 hidden nodes can achieve an extremely low SER at 15~dBm, performing significantly better than the D-DNN.

\begin{table}[ht]
    \centering
    \caption{Comparison the number of parameters for different network}
    \begin{tabular}{c c}
        \hline
        \textbf{Network} &\textbf{Number of Parameters (\textbf{Reducing rate})}  \\
        \hline
        D-DNN &398\\
        10 hidden nodes & 13728 (0\%) \\
      
        8 hidden nodes &11088 (19.2\%) \\
       
         6 hidden nodes &8488 (38.2\%) \\
          
         4 hidden nodes &5808 (57.7\%) \\
       
        2 hidden nodes & 3168 (76.9\%) \\
     
        Parameter sharing &468 (96.6\%) \\
      
         Removing $\mathcal{NN}_2$  & 416 (96.9\%) \\
         \textbf{parameter sharing ($N_h=8$)}& \textbf{378} (\textbf{97.2\%})\\
         \textbf{removing $\mathcal{NN}_2$ ($N_h=8$)}& \textbf{336} (\textbf{97.6\%})\\
         
        \hline
    \end{tabular}
    \label{tab:number of parameter}
\end{table}

\begin{table}[ht]
    \centering
        \caption{Comparison running time for different network}
    \begin{tabular}{c c}
        \hline
        \textbf{Network} &\textbf{Running Time (s) } \\
        \hline
         D-DNN & 7.03 \\
        10 hidden nodes & 13.01\\
      
        8 hidden nodes & 13.42 \\
       
         6 hidden nodes & 13.50 \\
          
         4 hidden nodes &13.60 \\
       
        2 hidden nodes & 13.61 \\
     
         Parameter sharing & 7.91 \\
      
         Removing $\mathcal{NN}_2$ & 3.85 \\
         \textbf{parameter sharing ($N_h=8$) } & \textbf{7.83}\\
          \textbf{Removing $\mathcal{NN}_2$ ($N_h=8$) } & \textbf{3.63}\\
         
        \hline
    \end{tabular}

    \label{tab:running time}
\end{table}

\subsection{Computational Complexity}
The running time of the aforementioned methods is compared in Table \ref{tab:running time}. Particularly, the methods with different numbers of hidden layer nodes have nearly identical runtime due to the unchanged depth of DNN. 
The two combined methods, parameter sharing in $\mathcal{NN}_2$ with 8 hidden nodes and removing $\mathcal{NN}_2$ with 8 hidden nodes, can significantly reduce runtime, cutting it by 5.18 seconds and 9.38 seconds, respectively. The former reduces the memory storage requirements for weights by using the same parameters, which alleviates the pressure on memory bandwidth and cache. The latter reduces runtime by decreasing the depth of the neural network. Moreover, the computational complexity of these two combined methods is $\mathcal{O}(N_h(L_t+N_t+L_r))$ and  $\mathcal{O}(N_h(L_t+L_r))$ which both increase linearly with $L_t$ and $L_r$.

\section{Conclusion}\label{VIII}
In this paper, we have evaluated the hardware imperfections in the THz UM-MIMO system and proposed a two-stage hardware imperfection compensation algorithm. In the first stage, we develop a DNN to represent the combined hardware imperfections. In the second stage, we choose to design either a digital precoder or combiner with NN to effectively compensate for these imperfections. Additionally, to balance the compensation performance and network complexity, we propose three methods to slim the DNN in the first stage including, pruning, parameter sharing, and removing $\mathcal{NN}_2$ from DNN. By combining these slimming methods, we can achieve an additional reduction in computational complexity. When the transmit power or SNR exceeds 5~dBm and 7.95~dBm, respectively, we apply both pruning and parameter sharing in $\mathcal{NN}_2$. 
When the transmit power or SNR falls below 5~dBm and 7.95~dB, respectively, we use a combination of pruning and removal of $\mathcal{NN}_2$. Extensive simulations have been conducted in both uncoded and coded systems to analyze the hardware imperfections and the performance of the proposed compensation algorithm and slimming methods. Numerical results show that the Tx compensation can perform better than the Rx compensation. Additionally, using the combined slimming methods can reduce parameters by 97.2\% and running time by 39.2\% while maintaining nearly the same performance in both uncoded and coded systems.

\input{ref.bbl}

\end{document}